# Evaluating the potential impacts of grey seal predation and fishery bycatch/discards on cod productivity on the Western Scotian Shelf and in the Bay of Fundy

Steven P. Rossi [a], Yanjun Wang [b], Cornelia E. den Heyer [c], Hugues P. Benoît [d]

[a] Simon Fraser University, Burnaby, BC, Canada

[b] St. Andrews Biological Station Fisheries and Oceans Canada, St. Andrews, NB, Canada

[c] Bedford Institute of Oceanography, Fisheries and Oceans Canada, Dartmouth, NS, Canada

[d] Institut Maurice Lamontagne, Fisheries and Oceans Canada, Mont-Joli, QC, Canada

# Abstract

The recovery of many groundfish stocks throughout the Northwest Atlantic has been impeded by elevated natural (i.e., non-fishing) mortality ($M$) among older/larger individuals. The causes of elevated mortality are not well known, though predation by rapidly growing grey seal herds and unreported fishing are thought to be possible drivers of mortality for Atlantic Cod (*Gadus morhua*) on the Western Scotian Shelf and in the Bay of Fundy (known as "4X5Y cod") and in nearby ecosystems. We developed a statistical catch-at-age model for 4X5Y cod that accounted for both grey seal predation and estimated bycatch/discards to evaluate the degree to which either of these factors may influence cod mortality. The model was fit over a range of predation and discarding scenarios to account for uncertainties and a lack of data for these processes. We found that most cod $M$ remained unaccounted for unless cod comprised a large proportion (>0.45) of the grey seal diet by weight. If the reported bycatch estimates are taken as accurate, then the magnitude of cod discards from non-directed fisheries was minor, though these estimates are highly uncertain.

# 1. Introduction

Groundfish populations in the Northwest Atlantic Ocean were fished to low abundance in the late 1980s and early 1990s (Myers et al. 1996, Sinclair and Murawski 1997) and most have not recovered despite widespread restrictions on fishing effort (Hilborn et al. 2021). Elevated rates of mortality arising from sources other than reported fishing (i.e., natural mortality or $M$) among older/larger individuals have been identified as the primary impediments to recovery for many groundfish stocks throughout the Northwest Atlantic (e.g., Swain & Mohn 2012, Swain et al. 2013, Swain and Benoît 2015, Rossi et al. 2019, Wang and Irvine 2022). The $M$ rates that some groundfish stocks are now experiencing are unsustainably high, more akin to those of small, short-lived forage fish than large, long-lived groundfish, placing some stocks at risk for extirpation in the coming decades (COSEWIC 2005, Swain and Chouinard 2008, Swain et al., 2019, Neuenhoff et al., 2019).

The drivers of groundfish $M$ in the Northwest Atlantic are difficult to establish and vary to some degree by species and ecosystem (Wiedenmann and Legault 2022). For instance, high rates of adult $M$ of Atlantic Cod (*Gadus morhua*) off Newfoundland and Labrador may be linked to capelin (*Mallotus villosus*) availability (Rose and O'Driscoll 2002, Regular et al., 2022). In other cases, the strong spatiotemporal correlation between groundfish $M$ and seal abundance has led many to hypothesize that increased predation by recovering pinniped populations is a driver of elevated groundfish $M$ (Mohn and Bowen 1996, Fu et al. 2001, Chouinard et al. 2005, Chassot et al. 2009,). This hypothesis has further been supported by other lines of evidence including a matching effect size, predation risk-related species distribution shifts and weight-of-evidence evaluation of other hypotheses (Benoît et al. 2011, Swain et al. 2011, 2015). Increases in $M$ may also alias emigration to adjacent ecosystems caused by other drivers like changes in the thermal environment, unreported catch and discarding (Wang and Irvine 2022, Wiedenmann and Legault 2022).

Hypotheses about sources of $M$ are difficult to evaluate, even for regularly assessed, data-rich species, such as those with age-structured information from both the fishery and surveys. Existing assessment models can be expanded to account for specific sources of mortality (e.g., mortality can be linked to predator information, or discards can be estimated by directly modelling fleets that capture the species-of-interest as bycatch), however, making reliable inferences from these models requires that they are fitted to data that are informative about the processes being evaluated. For instance, estimating the mortality imposed by a predator requires data on the abundance of predator populations, the spatiotemporal overlap between the predator and target species, and knowledge of predator diets. Similarly, estimating the implications of bycatch mortality requires accurate monitoring of cod catch and discards from those fleets and knowledge about the survival of discarded fish. This information is often vaguely known or entirely unavailable.

One example of a Northwest Atlantic groundfish stock with elevated $M$ due to uncertain factors is Atlantic Cod on the Western Scotian Shelf and in the Bay of Fundy (NAFO division 4X and the Canadian portion of 5Y; Figure 1), collectively managed as "4X5Y cod". This stock historically supported significant directed fisheries, averaging landings of nearly 25,000 mt annually from the mid-1960s to the early 1990s. However, cod biomass in 4X5Y declined to low levels in the late 1980s and early 1990s and has since continued to decline despite low fishing quotas. A number of factors may contribute to elevated cod $M$ in 4X5Y, including predation, unreported catch, and emigration to deeper or adjacent waters (Wang and Irving 2022). Studies have indicated that grey seal (*Halichoerus grypus*) predation has contributed to elevated $M$ in adjacent Northwest Atlantic cod stocks (e.g., Trzcinski et al. 2006; O'Boyle & Sinclair 2012, Neuenhoff et al. 2019), but the impact of grey seals on 4X5Y cod mortality has received less attention (Trzcinski et al. 2009). Cod interact with groundfish directed fisheries using mobile or fixed bottom gear and may also be incidentally captured in lobster traps. Cod may also be illegally retained as bait in fixed gear and lobster fisheries. Additionally, at-sea monitoring indicated that cod may be illegally discarded due to quota limitation, trip limits, or license restrictions. Discard estimates from the groundfish and lobster fisheries rely on at-sea observations that are extrapolated to reflect the entire fishery; however, observer coverage in both fisheries in most years are either absent, too low or not spatially representative, and so discards/bycatch estimates are considered highly uncertain (Gavaris et al., 2010, Pezzack et al., 2014, Clark et al., 2015). Moreover, the survival rate of discarded cod is unknown. There are no observations to support the hypothesis that large scale of emigration of 4X5Y cod drives $M$ increases despite comprehensive and frequent surveys and active fisheries in 4X5Y and adjacent management units in Canada and the US.

We investigated the potential role of grey seal predation and bycatch/discarding on 4X5Y cod mortality. We developed an age-structured model for 4X5Y cod that explicitly accounted for grey seal predation by including data about local seal presence, the bioenergetic requirements of grey seals, assumptions about grey seal diet composition, and information about discards. We used the model to estimate historical trends in abundance and mortality for 4X5Y cod and to project the relative impacts of predation and fishing on future stock abundance. Though data limitations precluded us from making direct inferences about the proportion of mortality attributed to each source, we instead aimed to make broad determinations about the plausibility of each source as a major driver of $M$ by evaluating models over a range of plausible scenarios.

## 2. Methods

**2.1. Data**

*2.1.1. Commercial fishery and survey*

Cod in 4X5Y are caught in multi-fleet commercial fisheries using both fixed (primarily in June-December) and mobile gear (year-round). Annual commercial landings were between 16-33 kt in the 1970s and 1980s but have steadily declined since the early 1990s to less than 1 kt in recent years (Figure A1). Most cod landings in 4X5Y prior to 1995 came from the Scotian Shelf, while Bay of Fundy landings were more predominant between 1995 and 2009. Landings have been evenly split between the two regions since 2010.

Cod biomass has been monitored since 1970 by the DFO-Maritimes Region summer research vessel survey (hereafter "RV survey"), which samples the Scotian Shelf and Bay of Fundy each July via a depth-area stratified design using bottom-trawl gear. Survey data comprise sampling by three different survey vessels: the A.T. Cameron (1970-1981; Yankee 36 trawl), the Lady Hammond (1982; Western IIa trawl) and the Alfred Needler (1983-2021; Western IIa). Biomass indices from each vessel are not directly comparable due to an absence of reliable conversion factors. In contrast, biological data such as weight-at-age are considered independent of vessel effects and are thus comparable across vessels. Needler biomass indices declined by approximately 86% from the 1980s to 2010s (Figure A2).

Age-composition (Figure A3) and weight-at-age of commercial and survey catches were available for 1970-2021. We note that the commercial samples for 1970-1982 may be biased as they are exclusively from the Scotian Shelf, and cod landed by commercial fisheries on the Scotian Shelf tended to be slightly younger than cod landed in the Bay of Fundy, while weight-at-age was generally smaller across all ages on the Scotian Shelf.

Cod are not sampled for maturity during the summer survey for logistical reasons, so maturity was inferred from Spring RV survey sampling that sporadically covers the Scotian Shelf and Bay of Fundy. Maturity-at-age curves have previously been estimated both on the Scotian Shelf and in the Bay of Fundy, though only in pre-2000 and post-2000 time blocks, due to the intermittent nature of the data (Andrushchenko et al., 2022). Bay of Fundy cod matured earlier than Scotian Shelf cod in each time block, while maturity occurring earlier in each area in the post-2000 time block. We calculated a maturity schedule for all of 4X5Y as the average maturity-at-age in each region, weighted by the relative abundance-at-age in each region from the RV survey (Figure A4). We did not include biomass indices or age-composition from the Spring RV survey in our analysis given the inconsistent nature of the survey.

Annual discards of cod by the groundfish fishery were estimated from at-sea observations that were scaled to the total catch of each fishery by weight (Gavaris et al. 2010, Clark et al. 2015) for 2003-2017 (mean: 22 t, range: 0-128 t). At-sea observers also collected data on bycatch from lobster (*Homarus americanus*) fisheries in 4X5Y, which was then used to estimate bycatch of each species by weight. A generalized additive model fitted to at-sea observations of bycatch in the lobster fishery for 1989-2018 estimated that annual discards ranged between 72 t and 460 t across this period (A. Cook, Fisheries and Oceans Canada, Dartmouth Canada unpublished analysis). A bycatch monitoring pilot program with a target coverage of 1% fishing trips was implemented in 2019-2021, a generalized linear model fitted to the 3-year combined data estimated the annual cod bycatch as 209 t (95% CI: 295-645 t) over this 3-year period (Cook et al. 2023). Both analyses implicitly assume that the magnitude and species-composition of bycatch was stationary over the periods analyzed. We note that discard estimates from both groundfish-directed fisheries (in all years) and lobster fisheries (1989-2018) may be unreliable due to very low levels of at-sea observer coverage across years, seasons, and gears for many of the principal fisheries and low cod abundance, while lobster discard estimates for 2019-2021 do not cover some areas such as the northern Bay of Fundy. We therefore only included bycatch/discard estimates in sensitivity analyses. The annual age-composition of cod discarded from the lobster fishery in 4X5Y was derived using the combined length frequency from bycatch sampling and year-specific age-length keys from the summer RV survey (Table A1).

Further details on fishery landings and survey data are given in Andrushchenko et al. 2018.

### 2.1.2. Grey seal presence in 4X5Y

Grey seals forage off the coast of North America from the northern tip of Labrador to the Northeast United States, with major breeding colonies in Canadian waters occurring at Sable Island, in the Gulf of St. Lawrence, and along the coast of Nova Scotia (Figure 1), and recently re-established colonies in the US (den Heyer et al. 2021). The proportion of time that grey seals spent foraging in 4X5Y (hereafter "the foraging rate") was estimated from the sex- and herd-specific movements of 208 grey seals tracked via satellite telemetry, deployed between 1993 and 2005 (Breed et al. 2006; Harvey et al., 2008; Swain et al., 2015; Table A2). No seal occurrences in the Canadian portion of NAFO area 5Yb were detected, so foraging rates in 4X were assumed to apply to 4X5Y as a whole. Foraging rates for seals from coastal Nova Scotia colonies could not be estimated due to a lack of data. We assumed that seals from colonies along the southwest coast of Nova Scotia (SWNS; Green Island, Round Island, Noddy Island, Mud Island, and Flat Island) foraged exclusively in 4X5Y given the proximity of these colonies to 4X5Y. Hay Island seals were assumed to have equivalent foraging behaviour to Gulf of St. Lawrence seals based on proximity and satellite tracking data suggesting that seals hauling out in this area are likely to forage in the Gulf of St. Lawrence (K. Whoriskey, pers. comm., 2019). Seals from the remaining colonies along Eastern Nova Scotia (ENS; Red Island, Basque Islands, White Island, Camp Island, Bowen's Ledge) were assumed to forage as Sable seals.

### 2.1.3. Grey seal abundance

A grey seal population dynamics model fitted to pup production, reproductive, and mark-resighting data from 1960-2021 was used to estimate age- and sex-specific grey seal abundance for the Gulf of St. Lawrence colonies and for a "Scotian Shelf" herd that was the aggregate of the Sable Island, ENS, and SWNS colonies (Rossi et al. 2021; Hammill et al. 2023). To determine abundance at SWNS colonies specifically, we first derived a time-series of SWNS pup production by fitting a logistic model to four observations of pup production in SWNS colonies, then calculated total SWNS abundance using the equilibrium ratio of pups to total abundance estimated from the population dynamics model (0.26:1; Appendix B).

Abundance estimates for the Gulf, SWNS, and Sable Island and ENS populations were multiplied by foraging rates to estimate grey seal foraging effort (seal-yrs) in 4X5Y.

### 2.1.4. Grey seal diet

The grey seal diet has not been formally studied in 4X5Y, so we used the global literature to define a range of plausible scenarios for the importance of cod to the grey seal diet. Grey seal diet studies analyzed gastrointestinal, scat, and/or fatty acid samples from 10,804 grey seals captured between 1950 and 2013 in the North Atlantic Ocean (Table A3). Reported proportions of cod in the grey seal diet by weight ranged from less than 0.01 to 0.68, with a mean of 0.15 (Figure A5). We regressed the proportional contribution of cod to the seal diet on survey timing, survey location (east or west Atlantic), sample size, and method, but found no significant relationship between these factors and cod consumption (Table A4).

## 2.2. Statistical catch-at-age model

We described the dynamics of 4X5Y cod using two statistical catch-at-age (SCA) models. The first model, termed the "RW" model, allowed $M$ to vary as a random walk. This model has previously been used to assess cod in the southern Gulf of St. Lawrence (Swain et al. 2015) and to estimate nonstationarities in cod and Yellowtail Flounder (*Limanda ferruginea*) populations on Georges Bank (Rossi et al. 2019). We also modified the RW model so that cod mortality was an explicit function of seal foraging effort and a hyperbolic (Type II) functional response (the "FR" model). The notation and equations for the models are presented in Table 1 and Table 2, respectively. In the following section, model equations are referred to in parentheses using the format (M.$X$) where $X$ refers to a specific model equation.

### 2.2.1. Population dynamics

The SCA operated on a yearly time-step, from 1970 to 2021, and from ages 1 to 7, with the oldest age-class accumulating fish as a plus-group. Cod abundance was assumed to decay exponentially according to a total instantaneous mortality rate $Z$ (M.13), which was partitioned into fishing mortality ($F$) and natural mortality ($M$) (M.3). $F$ was modelled as a separable function of an annual effect ($F_t$) and an age-specific selectivity ($S_{l,a}$) (M.1; details below). $M$ was further partitioned into grey seal predation mortality ($M^{(p)}$) and other natural mortality ($M^{(o)}$), where $M^{(o)}$ represented all $M$ not explicitly accounted for in $M^{(p)}$ (M.4). We estimated $M^{(o)}$ in three age blocks: (i) ages 1-2, (ii) ages 3-4 and (iii) ages 5-7+. $M^{(o)}$ varied as a random walk in time for the two oldest age blocks (M.7-M.8). We estimated a time-invariant value of $M^{(o)}$ for the youngest age block rather than estimate $M^{(o)}$ as a random walk, as correlations between recruitment and the mortality of young fish may cause estimates of $M^{(o)}$ for the youngest age block to be unreliable. $M^{(p)}$ was fixed at 0 in the RW model but varied as a random walk in the FR model. We calculated $F$ by solving the Baranov catch equation (M.18) using ten iterations of the Newton-Raphson method. This approach assumes that fishery catches were known exactly, and any mortality arising from unreported landings will be absorbed into $M^{(o)}$.

Fishery and survey selectivity were modelled as logistic, time-invariant functions of age (M.1). We explored two approaches for modelling time-varying fishery selectivity: (i) estimating selectivity in three time blocks (1970-1982, 1983-1994, 1995-2017), and (ii) varying selectivity parameters as random walks in time. The three time blocks were based on potential changes to fishery selectivity due to the possibility of biased estimate of fishery catch at age in 1970-1982 when the fishery sampling excluded the Bay of Fundy and the shift of fishing effort from Scotian Shelf to Bay of Fundy after 1994. The first approach yielded selectivity curves that were nearly identical, while the second approach created issues with model convergence. We therefore did not proceed with either approach to time-varying selectivity.

We initialized the model from a fished equilibrium (M.10-M.11). Annual recruit (age-1) abundance was assumed to follow a temporal random walk in log space (M.9, M.12).

In the FR model, grey seals were assumed to prey on all ages of cod according to an age-selectivity function (M.2). The age/size-selectivity of 4X5Y cod to grey seal predation is unknown due to a lack of diet studies. Diet studies from nearby ecosystems generally observed relatively few older cod in the seal diet (Benoit and Bowen 1990, Bowen et al. 1993, Bowen and Harrison 1994, Hammill et al. 2007); however, these data may simply reflect the age-composition of available cod rather than preferential predation on smaller/younger cod. This is supported by the results of Hammill et al. (2014), who found high percentages of large cod in grey seal diets when seals were sampled in areas where larger cod were aggregated. Whereas other studies may be biased towards smaller cod due to the locations from which grey seals were sampled, typically from shallow areas that may constitute or be near cod nursery areas. Furthermore, small cod could be more frequent in samples due to partial consumption (e.g., belly-biting) of larger cod such that prey hard parts are not observed (Benoît et al., 2011; Fallows et al. 2014). We assumed that the age-selectivity of cod to grey seal predation was asymptotic (i.e., older cod were preferred) by setting selectivity-at-age equal to proportion of cod mature-at-age. We also tested an alternative scenario in which younger cod were more fully selected (see section 2.2.4). The total number of cod that were vulnerable to seal predation was the product of cod abundance and seal selectivity, summed over ages (M.16). We note that, while smaller cod are often more highly prevalent than larger cod in grey seal diet samples, this may simply reflect the unavailability of large cod due to their low abundance rather than preferential selection.

The FR model contained a hyperbolic functional response of grey seals preying on cod that was based on the functional response used in Neuenhoff et al. (2019) for cod/grey seal interactions in the southern Gulf of St. Lawrence (M.17). The functional response assumed that the biomass of prey other than cod remained constant over all model years, an assumption that appears reasonable given the generalist nature of grey seals. Functional response parameters included the encounter rate between grey seals and cod ($f$), the proportional contribution of cod by mass to the average grey seal diet in a reference year ($d$), cod biomass in a reference year ($B^*$; kt), the maximum annual consumption by grey seals ($C_{\max}$; kt), and the average mass of cod vulnerable to seal predation ($w_t$; kt). The encounter rate was estimated externally from the model (see next section). We set the reference year to the first model year, 1970, as seal abundance was relatively low and cod biomass was relatively high during this period, suggesting that cod consumption per seal was probably nearer to asymptotic levels. We set $B^*$ to the estimated vulnerable biomass of cod in 1970 from the RW model and $C_{\max}$ to 2 tonnes based on bioenergetic estimates by Benoît et al. (2011). The average mass of vulnerable cod was also calculated from RW model estimates. We fixed $d$ across a range of values reported in the global literature. Each FR model name is given a suffix to indicate the proportional contribution of cod to the grey seal diet (e.g., $d$ is set to 0.15 in model FR-15).

Cod consumption by grey seals can be calculated from both the Baranov catch equation (M.19) and from the hyperbolic functional response (M.17, M.20). The terms of M.17 and M.20 can also be re-arranged so that seal abundance is calculated as a function of the number of cod consumed by grey seals (M.23; details in Neuenhoff et al. 2019). We used $M^{(p)}$ to calculate seal consumption from the Baranov equation, then inserted this estimate into M.23 to predict seal abundance.

*2.2.2. Objective function and estimation*

Observed biomass indices were assumed to be lognormally distributed (L.1) while observed proportions-at-age were assumed to arise from logistic-normal distributions (L.2). Gear-specific variance parameters for each distribution were conditionally estimated within the model. Full equations for these likelihoods are given in Rossi et al. (2019). Seal abundance was also assumed to be lognormally distributed (L.3).

We specified a standard normal prior for log-scale annual recruitment process deviations to allow large interannual changes in recruitment (L.4). Normal priors were specified for initial natural mortality, centered on 0.5 yr$^{-1}$ for the youngest age group (SD: 0.1) and on 0.15 yr$^{-1}$ for the other age groups (SD: 0.025) (L.5-L.7). A standard normal prior was also specified for initial abundance deviations to allow for large deviations from the equilibrium age structure (L.8). Wide normal priors were specified for initial recruitment and the selectivity parameters, with the expectation that the data would be informative about these parameters so the priors would have relatively little influence (L.9-L.11).

Annual process errors for $M^{(o)}$ and $M^{(p)}$ were assumed to follow zero-mean normal distributions with variance $\tau^{(o)}$ and $\tau^{(p)}$, respectively (L.12-L.13). We set $\tau^{(o)} = \tau^{(p)} = 0.075^2$, which is meant to allow the random walks sufficient freedom to achieve large shifts over time but provides enough constraint to prevent the random walk from being overly noisy.

Similar to Neuenhoff et al. (2019), we found there was little information in the data to estimate the encounter rate *f*, so we set the *f* in the model using the following steps:

1. Calculate cod consumption per seal in 1970 ($c^*$; millions) by assuming that the per-capita prey consumption in 1970 was exactly $C_{max}$ and that the proportion of cod in the seal diet was $d$, i.e.,

$$c^* = \frac{C_{max}d}{w_{t=1970}}$$

2. Set $c^*$ equal to the right-hand side of M.17, with vulnerable biomass calculated from RW model abundance and seal selectivity in 1970.

3. Find the numerical solution for $f$ ($f^*$) from the equation in step 2.

4. Set the encounter rate in the model at $f^*$.

We implemented the model using the Template Model Builder package (TMB; Kristensen et al. 2016) within R version 4.1.0 (R Core Team, 2021). Posterior distributions for parameters and predictive distributions for unobserved quantities of interest were generated from a Hamiltonian Monte Carlo (HMC; Duane et al. 1987, Neal 2011) algorithm known as the no-U-turn sampler (NUTS; Hoffman and Gelman 2014) using the tmbstan R package (Kristensen 2018). We ran three chains for 3000 iterations each (target average proposal acceptance = 0.99, maximum treedepth = 15), discarding the first half of each chain as a warmup. We monitored convergence using the potential scale reduction factor on rank-normalized split chains ($\hat{R}$), i.e., the ratio of between-chain variance to within-chain variance, where $\hat{R} > 1.01$ indicates problems with convergence (Vehtari 2020). For each parameter we also examined the effective sample size (ESS) of the rank-normalized draws, where ESS represents the number of independent draws that contain an equal amount of information as the MCMC samples. While higher values of ESS are better, we considered a minimum ESS of 300 (100 per chain) to be acceptable. We also monitored HMC/NUTS specific diagnostics, including the number of divergent transitions, the number of transitions that saturated the maximum treedepth, and the Bayesian fraction of missing information (Betancourt, 2016).

*2.2.3. Projection*

The fitted FR models were used to project cod abundance to 2030 under three fishing quota scenarios; specifically, the commercial fishery was assumed to catch either the full quota (825 t) annually, half the quota annually, or nothing. For each posterior sample, we fitted a Beverton-Holt model to spawning biomass and recruitment estimates, then projected the model based on parameter estimates from that sample, the fitted stock-recruitment model, each harvest level, and random draws of maturity-at-age and weight-at-age from the last five years of the historical data. SSB at the end of the projection period was compared to the limit reference point (LRP) of 22,193 t for this stock (Wang and Irvine, 2022).

*2.2.4. Sensitivity analyses*

To test the sensitivity of *M* estimates to assumptions about discarding, we considered alternative scenarios in which we directly account for the effect of bycatch/discarding by groundfish and lobster fisheries for 4X5Y cod. Specifically, we tested scenarios in which either the lower or upper bound of annual lobster fishery discards was taken from the cod stock each year for 1989-2019. In each scenario, we added point estimates of cod discards by weight from the groundfish fishery (2003-2017) to the total fishery catch in 4X5Y. This approach to accounting for groundfish discards assumes that the age-selectivity of cod captured and discarded by the groundfish fishery is equal to the age-selectivity of the directed 4X5Y cod. To account for the lobster fishery selecting for smaller cod than the commercial cod fishery, we modeled the lobster fishery as a separate fleet with selectivity arising from a gamma function, which allows selectivity to be dome-shaped (i.e., small- or medium-size cod are preferentially selected over large cod). Model-predicted proportions-at-age of cod caught in the lobster fishery were fitted to the observed age-composition of cod discards (Table A1). For each catch scenario we considered sub-scenarios in which discard mortality was either 100% or 25%. The former value is intended to serve as an illustrative lower bound on discard survival whereas the latter value was reported for cod discarded from the Gulf of Maine lobster fishery (Sweezey et al., 2020).

We also tested the sensitivity of predation mortality estimates from each FR model to assumptions about the maximum annual consumption by grey seals and the encounter rate. We tested four alternative values for maximum annual seal consumption (1.50 t, 1.75 t, 2.25 t and 2.50 t). For the encounter rate, we scaled the input value $f^*$ for FR model by 100, 10, 1/10, and 1/100.

Finally, we tested an alternative seal selectivity scenario in which cod aged 2 and older were fully selected while recruits were partially (0.5) selected. The alternative selectivity function was used by O'Boyle & Sinclair (2012) to characterize asymptotic selectivity for grey seals feeding on cod in NAFO area 4VsW and was consistent with previous cod/seal analyses for that area (Mohn and Bowen 1996; Trzcinski et al. 2006; Trzcinski et al. 2009).

## 3. Results

### 3.1. Model fits

The RW and FR model fit the biomass indices and age-composition reasonably well (Figure A6-Figure A7). Models assuming a large proportional contribution of cod to the grey seal diet provided the closest fits to the Cameron biomass indices; however, fits to the Needler biomass indices and Commercial fishery age-composition worsened as the cod prevalence in the diet increased, particulary in recent years (Figure 3).

We did not detect problems with MCMC convergence for any model; the potential scale reduction statistic was less than 1.01 for each parameter and the effective samples sizes were greater than 100 per chain for each parameter (Table A5). For some models, a small number of post-warmup transitions diverged (~0.1% of transitions), however, these divergences were spread throughout well-explored areas of the parameter space and were not localized or on a boundary and were therefore not considered problematic.

### 3.2. Historical estimates

Each model estimated similar total mortality trends, with $Z$ for age 3-4 cod increasing from approximately 0.4 yr$^{-1}$ in the early 1970s to 0.8 yr$^{-1}$ by the late 2000s, while $Z$ for ages 5-7 rose from 0.6 yr$^{-1}$ in the 1970s to 1.5 yr$^{-1}$ in recent years (Figure 4). Reported fishing was a major source of mortality in all models in the early 1990s and continued to be an appreciable source of mortality until 2010, with $F$ approximately doubling between 2000 and 2008. The main difference between the models was in the partitioning of mortality between $M^{(o)}$ and $M^{(p)}$. For instance, $M^{(o)}$ rose sharply in FR-02 between 1980 and 2010, as total mortality was high but little of this mortality was attributed to grey seal predation (Figure 4). In contrast, $M^{(o)}$ in FR-45 rose more gradually between 1980 and 2021, as seal predation accounted for much of the mortality increase over this period.

Total consumption of cod by grey seals increased until the early 2000s, then declined for a decade as cod biomass declined to low levels (Figure A8). From 2001 to 2010, consumption under FR-45 was approximately equal to commercial fishery catch. Consumption fluctuated in recent years but was greater than fishery catch since 2011 in all FR models except FR-02. In each of the four FR models, a greater number of deaths (by weight) were attributable to $M^{(o)}$ than predation; the only exception was for FR-45 from 2016 onward, when the number of deaths from each source was approximately equal (Figure A9).

Each model estimated a strong increase in SSB between 1970 and 1978, and steady decline thereafter, apart from brief resurgences in the late 1980s and mid-1990s (Figure A10; top row). Estimated SSB has been below the LRP since 2005. Annual recruitment has also declined, from an average of 46 million in the 1970s to less than 8 million recent years (Figure A10; middle row). The recruitment rate declined from the early 1970s to early 2000s but increased from less than 0.2 recruits per kg of SSB to more than 1.5 between 2003 and 2011 (Figure A10; bottom row). Stock-recruitment relationships were similar across models (Figure A11).

Assuming the age-selectivity of cod to grey seal predation was equal to the proportion of cod mature-at-age, grey seals most commonly consumed age-3 cod, followed by age-4, age-2, and age-5 (Figure 5; left column). The age 3-4 block annually accounted for nearly 55% of consumed cod by numbers (45% by weight), while the age 5-7+ annually accounted for 25% of consumed cod by numbers (45% by weight). Grey seals consumed proportionally more younger cod and fewer older cod in the 2010s than in the 1980s, as fewer older cod were available in later years.

The strength of density-dependence in the estimated functional response was positively related to the proportion of cod in the grey seal diet; the predation mortality rate imposed on cod per seal was more than twice as high at peak cod abundance than at low abundance under FR-45, whereas in FR-02 the predation mortality rate was less than 20% higher at peak abundance than at low abundance (Figure 6).

### 3.3. Projections

Projected cod SSB continued to decline when the seal population was assumed to grow as projected by the seal model and current levels of $M^{(o)}$ were maintained (Figure 7). Projected SSB was slightly lower in scenarios where fishery catches were taken, but SSB tended to decline even under no catch. Projections were only slightly more optimistic when seal abundance was projected to stabilize at 2019 levels (Table 4); in general, projected SSB continued to decline. Uncertainty around projection estimates was higher for models that attributed little mortality to seal predation (i.e., FR-02, FR-15) as these models attributed more mortality to unexplained sources (Figure 7).

### 3.4. Sensitivity analyses

Terminal estimates of source-specific mortality and spawning biomass for all sensitivity runs are summarized in Table A6-A10, while time-series of relative differences between estimates from sensitivity runs and the base models are shown in Figure 8. One general result is that estimates of total mortality were robust to changes in model specification (Table A9); changes in $F$ or $M^{(p)}$ in the sensitivity runs were always balanced by changes to $M^{(o)}$ to produce a similar value of $Z$.

Model fits to the lobster discard age-composition did not exhibit any troubling residual patterns (Figure A12). The estimated age-selectivity function for the lobster fishery was strongly dome-shaped, peaking at ages 3-4 (Figure A13).

Compared to the base models, accounting for the lower bound of total bycatch estimates and assuming 100% discard mortality led to $F$ increases of 7% for 2010-2018 and 40% for 2019-2021 (Figure 8). Alternatively, accounting for the upper bound of bycatch estimates with 100% discard mortality led to $F$ increases of 50% for 2010-2018 and 94% for 2019-2021 (Figure 8). Including estimated bycatch generally did little to explain the large increases in mortality over recent decades; for FR-02, FR-15, and FR-30, most deaths by weight were attributable to $M^{(o)}$ throughout the time-series (Figure 9).

Increasing the maximum annual consumption by grey seals by 0.5 t resulted in 25% increase in $M^{(p)}$, while decreasing annual consumption by 0.5 t resulted in a 25% decrease in $M^{(p)}$ (Figure 8). Predation mortality estimates were sensitive to decreases, but not increases, in the encounter rate (Figure 8).

The alternative seal selectivity function had relatively minor impacts on estimates of spawning biomass, total seal catch, or total mortality (Figure 10). The main effect of changing the selectivity function was to shift predation mortality from older cod to younger cod (Figure 5; right column). A larger number of cod were consumed under the alternative selectivity function than the base function, but the smaller size of these cod meant that the total weight of cod consumed by grey seals was similar under both functions (Figure 10).

## 4. Discussion

We developed a population model for 4X5Y cod accounting for reported fishery catch, grey seal predation and discard mortality and analyzed this model over a range of seal predation and bycatch/discard scenarios to evaluate the possible contribution of each factor to cod mortality. While the data needed to characterize the specific relationship between cod mortality and either grey seal predation or bycatch are not available for this stock, our analysis suggests that most cod natural mortality in 4X5Y is unexplained by either seal predation over a wide range of plausible predation scenarios or bycatch over a range of bycatch scenarios bounded by existing estimates. Natural mortality for adult cod began rising rapidly in the mid-1980s and reached very high levels (0.75-0.95 yr$^{-1}$) by the late 1990s. Grey seal abundance was estimated to have been low (< 2000 seals) during this period, though there are no direct population counts from grey seal colonies along southwest Nova Scotia until 2006. Similarly, estimated bycatch was a small fraction of reported landings until recent years and was thus a very marginal factor in cod mortality, historically. Notably, for all scenarios in which cod comprised 30% of the grey seal diet by weight or less, the proportion of cod deaths attributable to mortality arising from sources other than fishing or grey seal predation ($M^{(o)}$) increased over time, suggesting that influential population processes were not being directly captured by the model or that mortality inputs from fishery sources are underestimated. The proportion of cod deaths attributable to $M^{(o)}$ was only relatively constant or non-increasing when cod comprised a very large (e.g., >45%) share of the grey seal diet. We note, however, that despite the relatively small historical influence of bycatch in all models and of seal predation in models where seals consumed modest amounts of cod, both bycatch/discarding and seal predation will increase in importance in coming years if cod biomass continues to decline.

Current estimates of cod bycatch/discards from the groundfish and lobster fishery suggest that bycatch/discarding represents a minor component of total mortality, however these estimates are impaired by low observer coverage in the groundfish fishery in most years and uncertainties about the survival of discarded fish. Cod bycatch and discarding could be missed entirely when observer coverage is low, especially in recent years with cod at very low abundance. If true discards are several times greater than current "upper bound" estimates then discard mortality could be significant source of mortality. There has also been a substantive increase in both lobster and Atlantic halibut landings in 4X5Y (Bernier et al. 2018), suggesting that there may be changes in the proportion of cod in the catch. While it is unknown how an increase in the target species abundance and/or fishing effort would impact cod bycatch, it should be noted that both these fisheries require bait. Monitoring the potential impact of these fisheries on 4X5Y cod requires maintaining recent observer coverage in the lobster fishery and increasing observer coverage in groundfish-directed fisheries, particularly as harvesters may be able to modify fishing practices to avoid or increase bycatch. We also note that reducing uncertainties around bycatch mortality may be more tractable than reducing uncertainty around predation mortality since the latter is a product of processes that are similarly unknown but more difficult to observe.

Functional response parameters were a major source of uncertainty in our analysis given the sparsity of grey seal data in this region. The model was predictably sensitive to assumptions about the proportion of cod in the grey seal diet. These parameters could not be reliably estimated in the model and needed to be inferred from external analyses and data from other ecosystems. Estimating grey seal diet composition is difficult as grey seals forage widely and are opportunistic feeders. In initial trials, we attempted to estimate $d$ within the model using two different priors, $U(0,1)$ and Beta(1.42,7.84), where the parameters for the latter distribution were generated by fitting a beta distribution to global estimates of $d$. However, under these formulations the model was extremely sensitive to the allocation of process error. For instance, when $\tau^{(p)} > \tau^{(o)}$, the estimate of $d$ was around 0.42 and model estimates were similar to FR-45. In contrast, when $\tau^{(p)}$ was equal to or less than $\tau^{(o)}$, then nearly all increases in $M$ were absorbed into $M^{(o)}$ and estimated $d$ was around 0.02. Models with fixed $d$ were more robust to *a priori* choices about $\tau^{(p)}$ and $\tau^{(o)}$.

The hyperbolic functional response in our analysis was borrowed from a model of seal/cod interactions in the southern Gulf of St. Lawrence (Neuenhoff et al., 2019) and had two main advantages. First, the functional response incorporated a bioenergetic estimate of the maximum annual prey consumption by grey seals, which bounded prey consumption in our model to a biologically plausible range. Second, the functional response could be rearranged and fit to estimates of local grey seal abundance. We initially attempted an alternative, more flexible approach to modelling cod/seal interactions by allowing per-capita consumption to vary as a random walk, then fitting a functional response to the resulting estimates of consumption of prey biomass (Cook et al. 2015). However, under this approach, predation mortality absorbed all of natural mortality, which lead to highly suspect behaviour, such as individual seals consuming several tonnes of cod annually. We therefore considered the more constrained approach of Neuenhoff et al. (2019) to be more suitable for modelling cod/seal dynamics in the data-limited 4X5Y area, as it was less prone to chase statistical noise. Cook et al. (2015) bounded consumption in their model by fitting to external estimates of prey consumption; however, comparable estimates are not currently available for 4X5Y cod. We did not attempt to fit a sigmoidal (Type III) functional response given the lack of predation data and we note that prey-switching at low cod abundance, which would be expected under a sigmoidal functional response, has not emerged in the southern Gulf of St. Lawrence despite the very low abundance of cod in that ecosystem.

The impact of grey seal predation on cod in NAFO area 4X was previously assessed by Trzcinski et al. (2009). Relying on the grey seal diet data from the Scotian Shelf, Trzcinski et al. assumed that cod comprised 2% of the grey seal diet by weight and that seals primarily selected younger cod for consumption They found that grey seal predation accounted for a minor amount of mortality, with grey seals consuming approximately 100,00-200,000 cod annually between 1985 and 2007, and that high $F$ (0.16 yr$^{-1}$ in 2007) and other sources of $M$ (0.66 yr$^{-1}$ in 2007) were the main drivers of the stock decline. In comparison, our FR-02 model estimated that grey seals typically consumed between 80,00-300,000 cod annually between 1985 and 2007, and that $F$ was 0.33 yr$^{-1}$ in 2007 and that other $M$ was for 0.55 yr$^{-1}$ for ages 1-2, 0.41 yr$^{-1}$ for ages 3-4, and 1.00

yr$^{-1}$ for ages 5-7+. Therefore, despite some differences in model structure, assumptions and estimates of non-predation mortality, our FR-02 consumption estimates are broadly similar to estimates from the functional response model from Trzcinski et al. (2009).

The shape of the selectivity-at-age of cod to grey seal predation is also a key uncertainty in this system. Many studies of cod/seal interactions assume a "dome-shaped" selectivity function that preferentially selects younger cod. In our analysis, we demonstrated that younger cod may still dominate seal diet composition even when older cod are fully selected. This suggests that reported grey seal diet compositions may be more reflective of prey availability than prey preference. "Belly-biting" (i.e., the partial consumption of the abdomen of larger cod) by grey seals, which has been reported by fish harvesters in adjacent ecosystems (e.g., O'Boyle and Sinclair, 2012; Neuenhoff et al., 2019), will further bias diet samples towards smaller cod. Our choice of an asymptotic selectivity function is supported by diet sampling in the Cabot Strait, which demonstrated that grey seals will eat large cod and potentially select for them when large cod are available and aggregated (Hammill et al., 2014).

Our analysis only included grey seals that whelp in Canadian waters, however, increasing numbers of grey seals are also pupping in the Gulf of Maine. U.S. breeding colonies produced at least 6,308 pups in 2016, suggesting a total population size of 27,131 if the pup to adult ratio for U.S. grey seals is consistent with Canadian grey seals (NOAA 2020). The degree to which U.S. grey seals forage in Canadian waters is unknown, because information about their movement is limited to one study that tagged 9 individuals on Cape Cod. Nevertheless, seals in this study did move considerable distances, with one nearing Sable Island (Puryear et al. 2016). It is therefore plausible that U.S. grey seals, particularly those pupping further north in the Gulf of Maine, could forage in 4X5Y. Our model may therefore underrepresent the true level of grey seal foraging effort in 4X5Y. Further research on U.S. grey seal movement and more robust estimates of total population size should improve the evaluation of the impact of these seals on Northwest Atlantic fish stocks.

Grey seals from colonies along coastal Nova Scotia accounted for about half of the estimated grey seal foraging effort in 4X5Y in 2019, and were projected to account for a majority by 2024 under all catch scenarios; however, the abundance and foraging behaviour of these seals are both highly uncertain. The dynamics of seals from coastal Nova Scotia colonies are difficult to model due to a lack of data and suspected immigration from other herds. Additionally, the foraging behaviour of SWNS seals in our analysis was based strictly on assumptions, as data were unavailable to estimate the proportion of time these seals spent foraging in 4X5Y. These levels of seal presence could have severe impacts for the primary prey species of grey seals on the Scotian Shelf, whether those prey are cod or another species, and suggest a need to better understand the dynamics of grey seals along coastal Nova Scotia.

## Acknowledgements

The work received financial support from a Fisheries Science and Ecosystem Research Program (FSERP) and the Deputy Minister (DM) Results Fund of Fisheries and Oceans Canada. We would like to thank Adam Cook for providing cod bycatch estimates from the lobster fishery, Irene Andrushchenko for providing 4X5Y cod landings and research survey data, and Fonya Irvine for assistance with updating cod discards from groundfish fishery. Adam Cook, Irene Andrushchenko and Sean Cox provided helpful feedback on earlier drafts of this paper.

# Tables

*Table 1. Notation for 4X5Y cod statistical catch-at-age model.*

| Symbol | Description | Value |
| --- | --- | --- |
| *Indices* | | |
| $A$ | Maximum age class (yr) | 7 |
| $K$ | Number of natural mortality age blocks | 3 |
| $t$ | Year | 1983,…,2017 |
| $a$ | Age class (yr) | 1,…,$A$ |
| $g$ | Fishing gear index | 1 (Commercial fishery), 2 (Survey, Cameron), 3 (Survey, Needler) |
| $k$ | Age-block index | 1,…,$K$ |
| *Data and inputs* | | |
| $C_t$ | Commercial fishery catch (kt) | |
| $I_{g,t}$ | Survey biomass index for gear $g$ in year $t$ ($g>1$) | |
| $u_{g,a,t}$ | Age-composition for gear $g$ in year $t$ | |
| $E_t$ | Grey seal foraging effort in 4X5Y (seal-yrs) | |
| $h_g$ | Timing of fishery $g$ (ordinal start-date of fishery / 365) | |
| $m_{a,t}$ | Maturity-at-age | |

| | |
|---|---|
| $w_{g,a,t}$ | Cod mass per unit of abundance in fishery/survey $g$ (kt) |
| $w_{a,t}^{(B)}$ | Beginning-of-year mass per unit of abundance (kt) |
| $w_t$ | Average mass per unit of cod abundance vulnerable to seal predation (kt) |
| $B^*$ | Cod biomass vulnerable to seal predation in 1983 (kt) |
| $C^*$ | Maximum annual grey seal consumption (kt), $C^* = 2.0$ |
| $d$ | Proportional contribution of cod by mass to the average seal diet in 1983 |
| $f$ | Encounter rate between grey seals and vulnerable cod |

*Parameters*

| | |
|---|---|
| $\bar{R}$ | Mean recruitment (millions) |
| $p^{(\text{init})}$ | Initial fully-selected predation rate (yr$^{-1}$) |
| $o_i^{(\text{init})}$ | Initial other natural mortality rate by age block (yr$^{-1}$) |
| $s_{g,t}^{(50\%)}$ | Age-at-50% selectivity |
| $s_{g,t}^{(95\%)}$ | Age-at-95% selectivity |
| $R^{(\text{init})}$ | Initial recruitment (millions) |
| $F^{(\text{init})}$ | Initial fishing mortality (yr$^{-1}$) |
| $\varepsilon_a^{(\text{init})}$ | Initial abundance-at-age deviations |
| $\varepsilon_t^{(P)}$ | Predation process error |

$\varepsilon_t^I$          Recruitment process errors

---

*Latent variables*

$N_{a,t}$          Abundance (millions)

$M_{a,t}$          Instantaneous natural mortality rate (yr$^{-1}$)

$F_t$          Instantaneous fully-selected fishing mortality rate (yr$^{-1}$)

$P_t$          Instantaneous fully-selected predation mortality rate (yr$^{-1}$)

$Z_{a,t}$          Instantaneous total mortality rate (yr$^{-1}$)

$m_{a,t}$          Proportion mature at age *a* in year *t*

$S_{g,a}$          Selectivity to gear *g* at age *a*

$C_a^{(p)}$          Number of cod eaten by seals (millions)

$B_t$          Spawning biomass at the start of year *t* (kt)

$V_t$          Number of cod vulnerable to predation in year *t* (millions)

$C_a^{(p)}$          Number of cod eaten by seals (millions)

$\hat{I}_{g,t}$          Predicted biomass index

$\hat{u}_{g,a,t}$          Predicted age-composition

$\hat{E}_t$          Predicted grey seal foraging effort in 4X5Y (seal-yrs)

*Table 2. Equations for 4X5Y cod statistical catch-at-age model.*

| Equation | Formula |
|---|---|
| **Age selectivity** | |
| (M.1) Fishery/survey selectivity | $S_{g,a} = \left(1 + \exp\left[\dfrac{-\ln(19)\left(a - s_g^{50\%}\right)}{s_{g,t}^{95\%} - s_g^{50\%}}\right]\right)^{-1}$ |
| (M.2) Seal selectivity | $S_{a,t}^{(p)} = m_{a,t}$ |
| **Mortality** | |
| (M.3) Total mortality | $Z_{a,t} = F_t\, S_{1,a} + M_{a,t}$ |
| (M.4) Natural mortality | $M_{a,t} = M_t^{(p)} S_{a,t}^{(p)} + M_{a,t}^{(o)}$ |
| (M.5) Initial predation mortality | $M_{1970}^{(p)} = p^{(\text{init})}$ |
| (M.6) Predation mortality ($t > 1970$) | $M_t^{(p)} = M_{t-1}^{(p)} \exp\left(\varepsilon_t^{(p)}\right)$ |
| (M.7) Initial other natural mortality | $M_{a,1970}^{(o)} = o_i^{(\text{init})},\ a \in k_i$ |
| (M.8) Other natural mortality ($t > 1970$) | $M_{a,t}^{(o)} = M_{a,t-1}^{(o)} \exp\left(\varepsilon_{k_i,t}^{(o)}\right),\ a \in k_i$ |
| **State dynamics** | |
| (M.9) Initial recruitment | $N_{1,1970} = R^{(\text{init})}$ |
| (M.10) Initial abundance, $1 < a < A$ | $N_{a,1970} = N_{a-1,1970} \exp(-F^{(\text{init})} S_{1,a-1} - M_{a-1,1970} + \varepsilon_a^{(\text{init})})$ |

(M.11) Initial plus-group abundance

$$N_{A,1970} = \frac{N_{A-1,1970}\exp(-M_{A-1,1970} + \varepsilon_a^{(init)})}{1 - \exp(-M_{A-1,1970} + \varepsilon_a^{(init)})}$$

(M.12) Recruitment, $t > 1970$

$$N_{1,t} = N_{1,t-1}\exp\left(\varepsilon_t^{(R)}\right)$$

(M.13) Abundance, $1 < a < A$, $t > 1970$

$$N_{a,t} = N_{a-1,t-1}\exp(-Z_{a-1,t-1})$$

(M.14) Plus-group abundance, $t > 1970$

$$N_{A,t} = \sum_{a=A-1}^{A} N_{a,t-1}\exp(-Z_{a,t-1})$$

(M.15) Spawning stock biomass

$$B_t = \sum_a N_{a,t}\, w_{a,t}^{(B)}\, m_{a,t}$$

(M.16) Cod vulnerable to predation

$$V_t = \sum_a N_{a,t}\, S_a^{(p)}$$

(M.17) Cod consumption per seal (numbers)

$$c_t = \frac{fV_t}{1 + f[B^*(d^{-1}-1) + V_t w_t]/C_{\max}}$$

(M.18) Fishery catch (weight)

$$C_t = \sum_a \frac{F_t\, S_{1,a}}{Z_{a,t}} N_{a,t} w_{1,a,t}(1 - \exp(-Z_{a,t}))$$

(M.19) Cod killed by seals (Calc. 1; numbers)

$$C_t^{(p)} = \sum_a \frac{M_t^{(p)} S_a^{(p)}}{Z_{a,t}} N_{a,t}(1 - \exp(-Z_{a,t}))$$

(M.20) Cod killed by seals (Calc. 2; numbers)

$$C_t^{(p)} = c_t E_t$$

**Observations**

(M.21) Biomass indices, $g>1$

$$\hat{I}_{g,t} = q_g \sum_a N_{a,t} S_{a,g,t}\, w_{g,a,t}\exp(-h_g Z_{a,t})$$

(M.22) Age-composition
$$\hat{u}_{g,a,t} = \frac{N_{a,t} S_{a,g,t} \exp(-h_g Z_{a,t})}{\sum_b N_{b,t} S_{b,g,t} \exp(-h_g Z_{b,t})}$$

(M.23) Seal foraging effort
$$\hat{E}_t = \frac{C_t^{\text{seal}}}{fV_t}\left(1 + \frac{f[B^*(d^{-1} - 1) + V_t w_t]}{C^*}\right)$$

*Table 3. Objective function components for 4X5Y cod statistical catch-at-age model.*

| Component | Distribution |
|---|---|
| **Likelihood** | |
| (L.1) Biomass indices | $\ln(I_{g,t}) \sim N(\ln(\hat{I}_{g,t}), \sigma^2_{I,g})$ |
| (L.2) Age-composition | $\boldsymbol{u}_{g,t} \sim P\left(N(\hat{\boldsymbol{u}}_{g,t}, \sigma^2_{u,g})\right)$ |
| (L.3) Seal effort | $\ln(E_t) \sim N(\ln(\hat{E}_t), \sigma^2_{E,t})$ |
| **Priors** | |
| (L.4) Recruitment deviations | $\varepsilon^R_t \sim N(0,1)$ |
| (L.5) Initial natural mortality, ages 1-2 | $M_1 \sim N(0.5, 0.1)$ |
| (L.6) Initial natural mortality, ages 3-4 | $M_2 \sim N(0.15, 0.025)$ |
| (L.7) Initial natural mortality, ages 5-7+ | $M_3 \sim N(0.15, 0.025)$ |
| (L.8) Initial abundance deviations | $\varepsilon^{(init)}_a \sim N(0,1)$ |
| (L.9) Age-at-50% selectivity | $s^{(50\%)}_{g,t} \sim N(2.5, 2)$ |
| (L.10) Selectivity step | $s^{(\Delta)}_{g,t} \sim N(1,1)$ |
| (L.11) Other natural mortality deviations | $\varepsilon^M_{k_i,t} \sim N(0, \tau^{(o)})$ |
| (L.12) Predation mortality deviations | $\varepsilon^P_t \sim N(0, \tau^{(p)})$ |

*Table 4. Proportion of projected posterior samples in which spawning biomass in 2030 exceeds the limit reference point for each FR model assuming that (i) projected seal effort is either fixed at 2021 levels or grows as projected by the grey seal population model, and (ii) age-selectivity of cod to grey seal predation is either equal to cod maturity-at-age (i.e., "base" selectivity) or equal to 0.5 for age-1 cod and equal to 1 for cod aged 2 or older (i.e., "full" selectivity).*

| Scenario | FR-02 | FR-15 | FR-30 | FR-45 |
|---|---|---|---|---|
| *Constant seal effort, base selectivity* | | | | |
| Full quota | 0.029 | 0.014 | 0.004 | 0.000 |
| Half quota | 0.034 | 0.018 | 0.007 | 0.000 |
| No catch | 0.040 | 0.016 | 0.008 | 0.001 |
| *Constant seal effort, alternative selectivity* | | | | |
| Full quota | 0.027 | 0.010 | 0.001 | 0.000 |
| Half quota | 0.031 | 0.013 | 0.001 | 0.000 |
| No catch | 0.037 | 0.015 | 0.002 | 0.000 |
| *Increasing seal effort, base selectivity* | | | | |
| Full quota | 0.024 | 0.013 | 0.003 | 0.000 |
| Half quota | 0.034 | 0.016 | 0.004 | 0.000 |
| No catch | 0.040 | 0.019 | 0.005 | 0.000 |
| *Increasing seal effort, alternative selectivity* | | | | |
| Full quota | 0.026 | 0.009 | 0.000 | 0.000 |
| Half quota | 0.031 | 0.011 | 0.000 | 0.000 |
| No catch | 0.037 | 0.022 | 0.001 | 0.000 |

# Figures

*Figure 1. Main grey seal pupping colonies in Canada (circles). Circle colours indicate whether the colony was grouped into the Gulf herd (red), Eastern Nova Scotia herd (blue), or Southwest Nova Scotia herd (green) for characterizing foraging behaviour. Orange lines delineate the Northwest Atlantic Fisheries Organization division boundaries. The cod stock in division 4X and the northeastern portion of division 5Y that lies within Canada's exclusive economic zone (EEZ) is collectively managed and referred to as "4X5Y cod".*

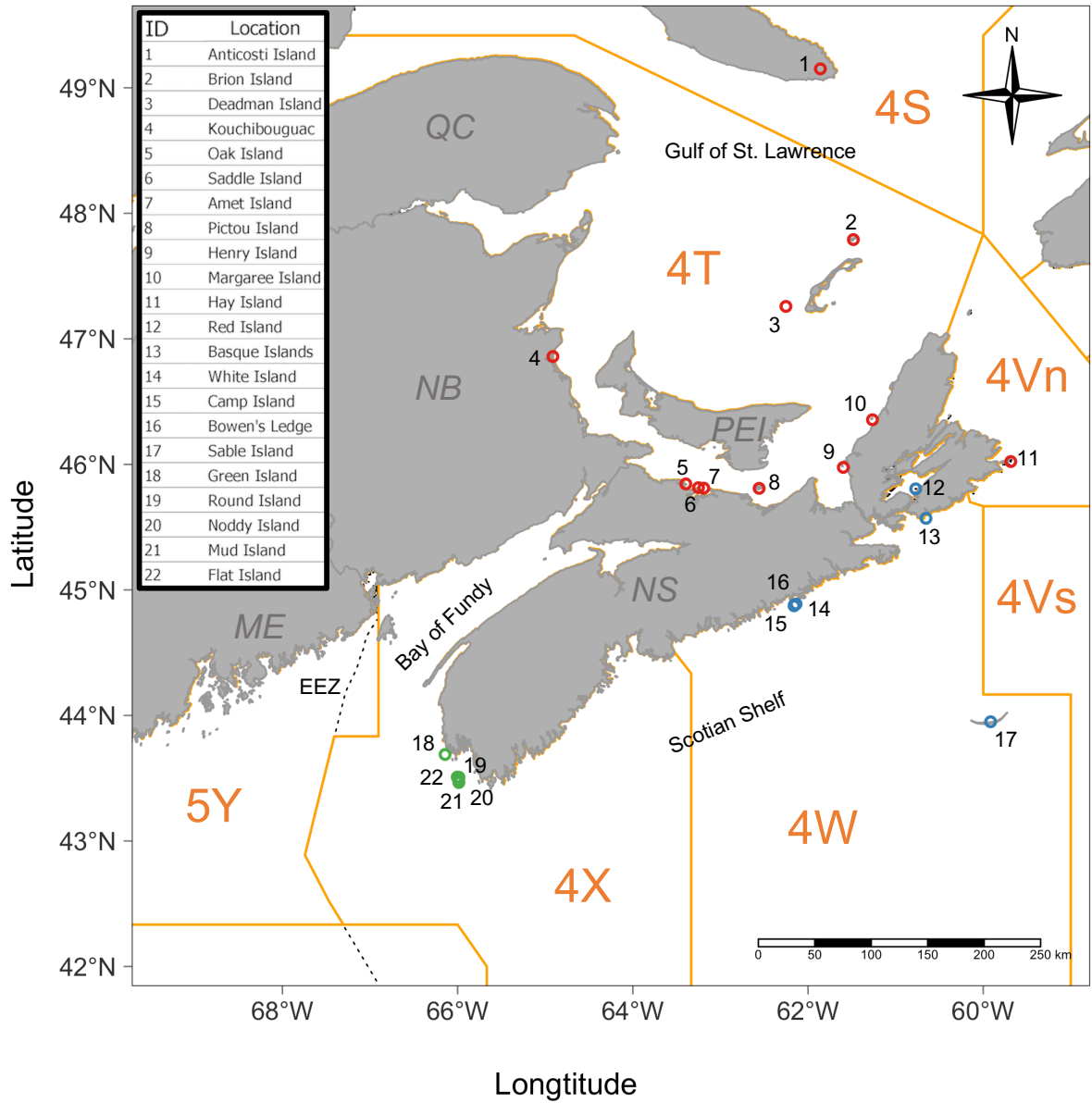

*Figure 2. Estimated grey seal foraging effort in NAFO Div. 4X for the total grey seal population and by subpopulation. Lines are posterior modes while shaded regions are central 95% uncertainty intervals. The black vertical line separates historical effort, which was estimated from data, and projected effort.*

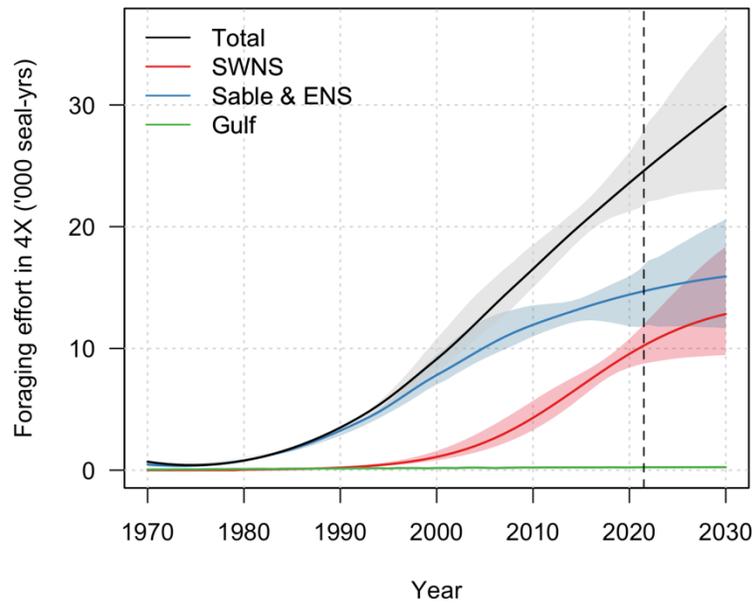

*Figure 3. Estimated residual variance for each model by data type (top row: biomass indices, bottom row: age-composition) and source (columns). Points are posterior modes, thick lines central 50% uncertainty intervals, and think lines are central 95% uncertainty intervals.*

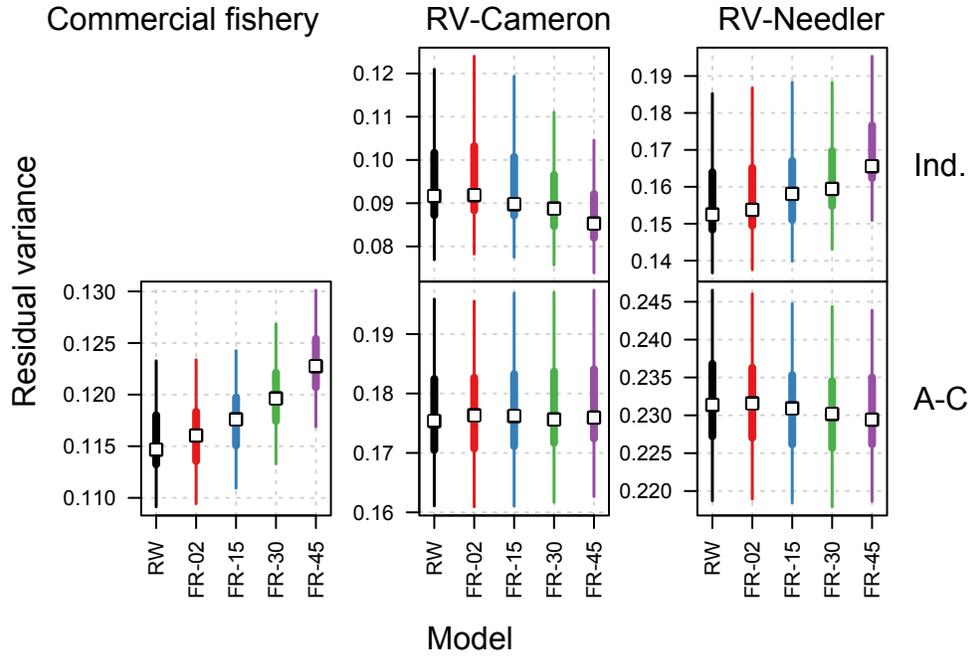

*Figure 4. Estimated mortality time-series (posterior modes) from five 4X5Y cod models.*

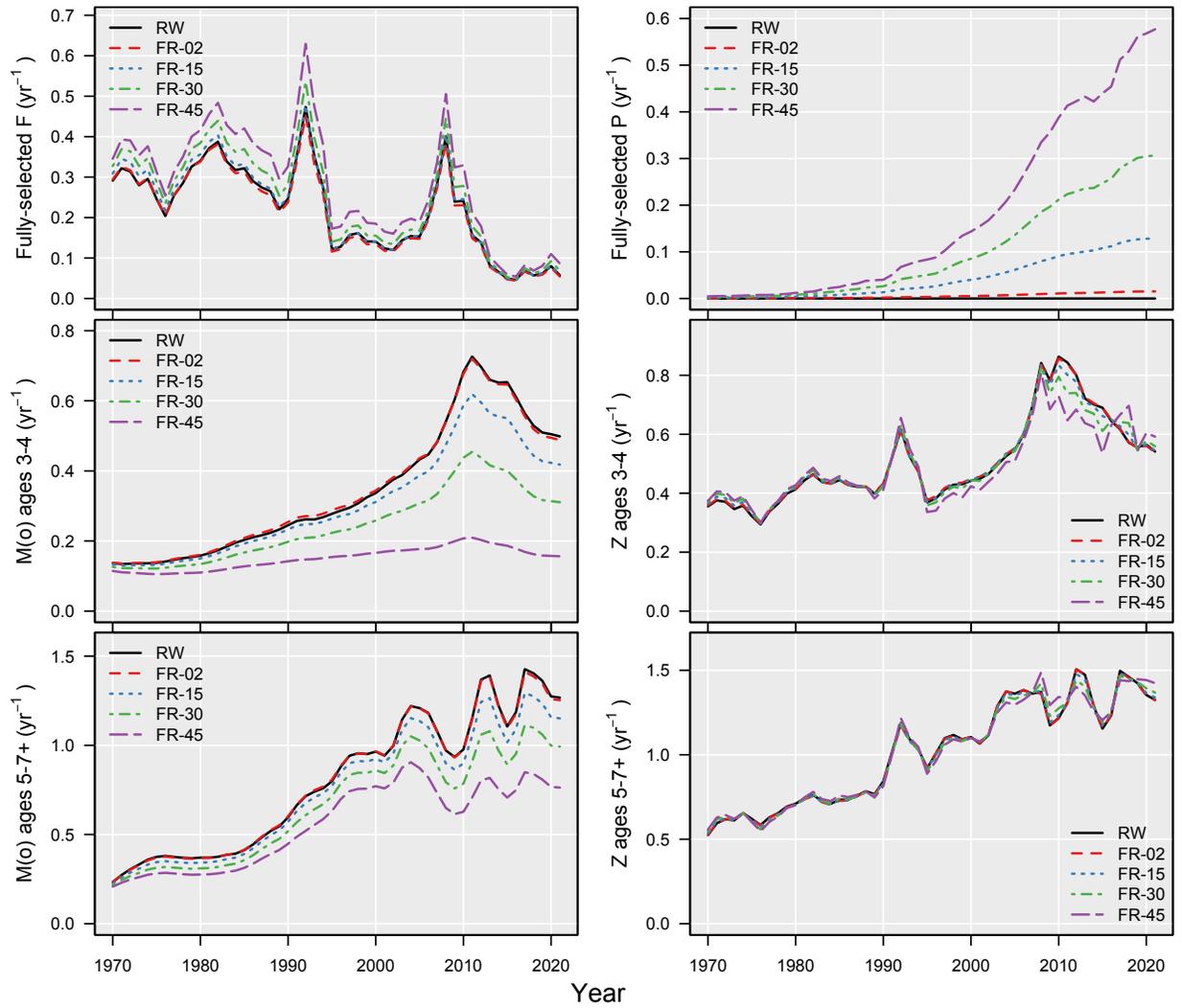

*Figure 5. Mean decadal consumption of cod by grey seals estimated from four FR models (rows), assuming that either the age-selectivity of cod to grey seal predation was equal to the proportion of cod mature-at-age (left column) or assuming all cod age 2 and older were fully-selected by grey seals with recruit selectivity equal to 0.5 (right column).*

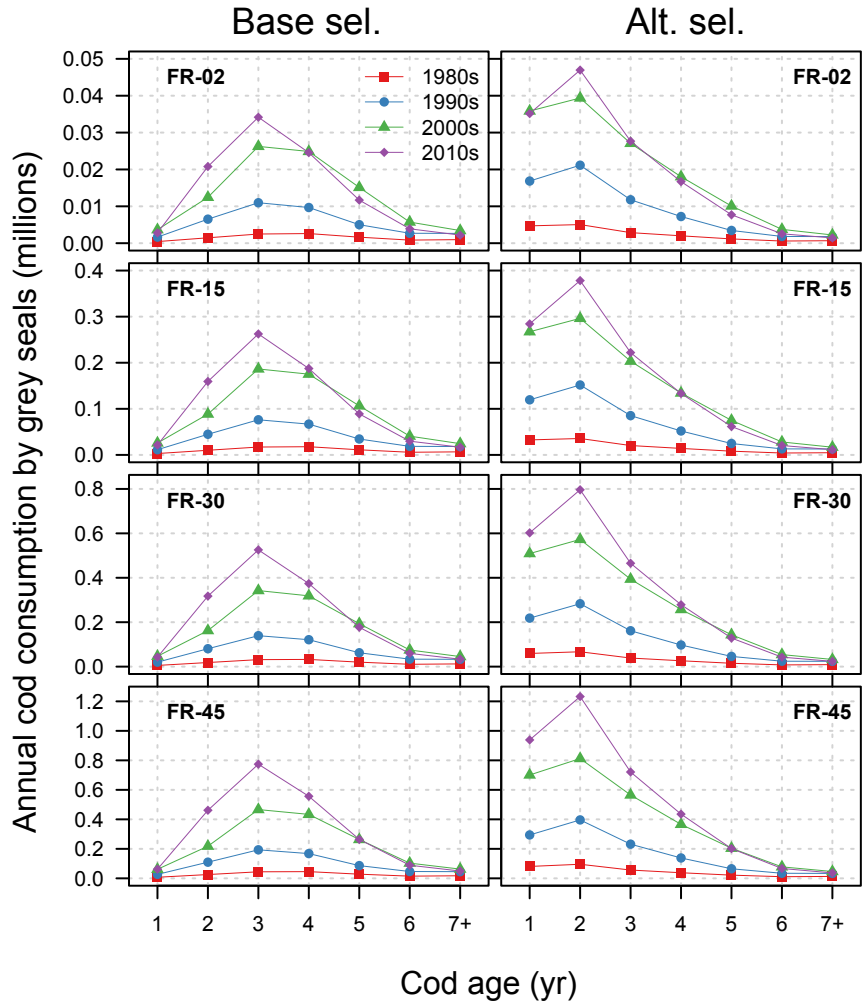

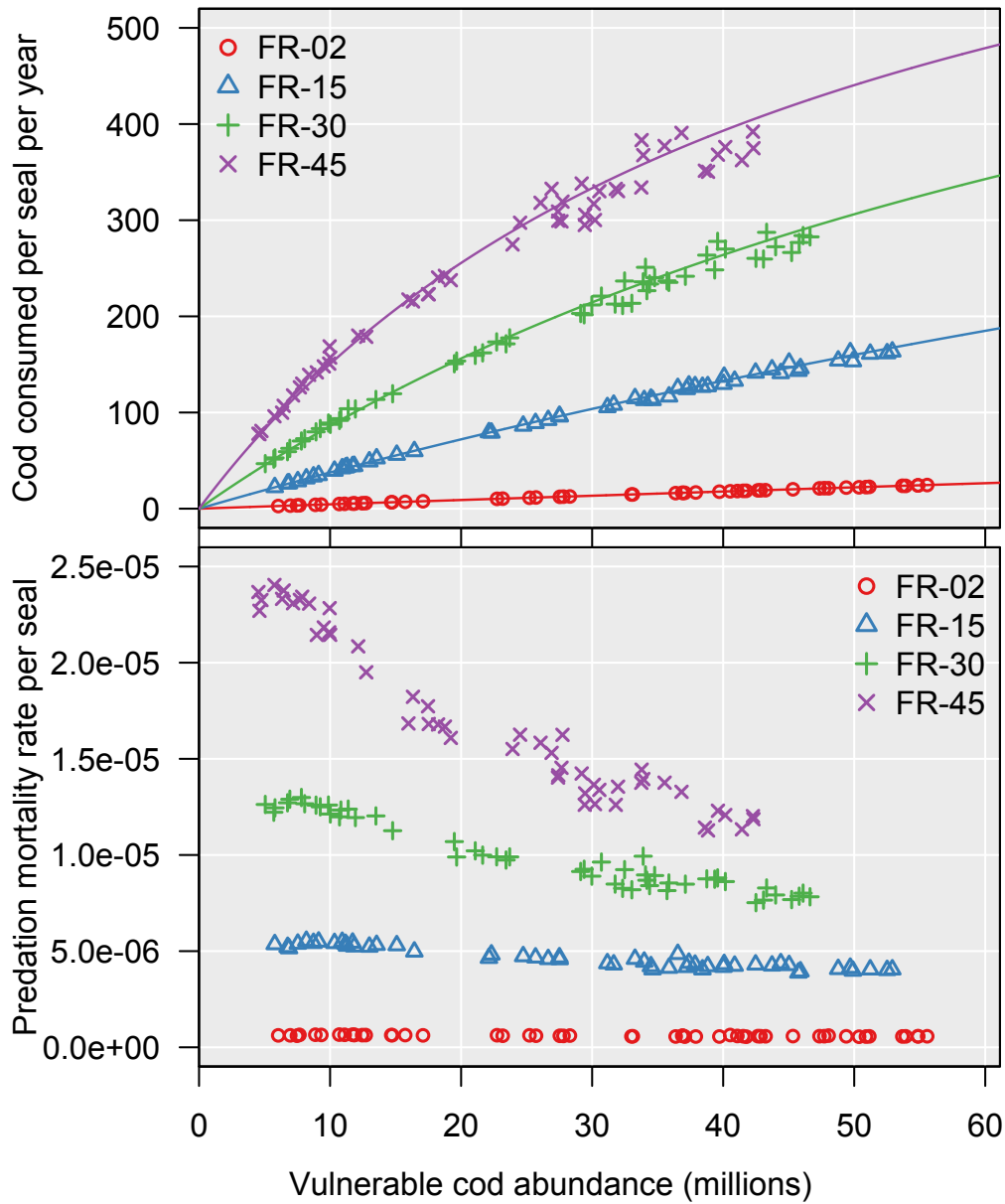

*Figure 6. Estimated functional response of grey seals preying on 4X5Y cod for five FR models (top row) and the resulting mortality rate per seal imposed on cod (bottom row). Points represent posterior modes of biomass and consumption for 1970-2021 using a functional response that accounted for interannual differences in cod mass. Lines represent the predicted functional response based on the mean mass of vulnerable cod for 1970-2021. Functional responses based on the mean mass of cod were used in the projections.*

*Figure 7. Projected spawning stock biomass (left column) and seal predation mortality rate (right column) for 4X5Y cod for four FR models (rows) assuming the grey seal population continues to increase between 2020-2029 as projected by the seal model. Three separate harvest strategies were assumed: fullQuota (825 t was caught annually), halfQuota (412.5 t was caught annually), and noCatch (no fishery catches).*

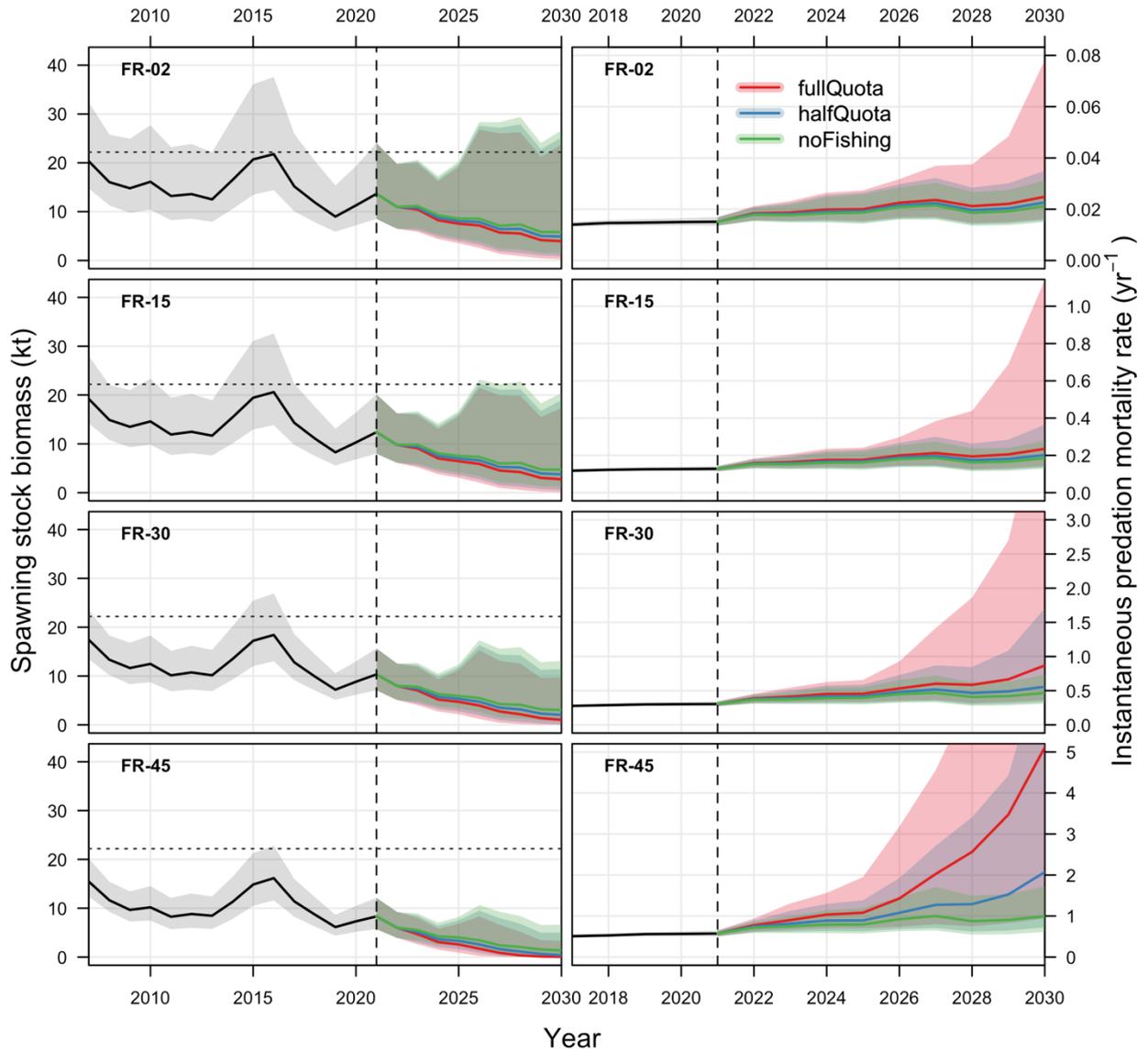

*Figure 8. Difference in cod mortality estimates between the Base model and six sensitivity runs. Difference is calculated as s-b where b is an estimate from the base model and s is a corresponding estimate from a sensitivity run. Sensitivity runs were defined by setting discards to the (a) lower and (b) upper bound of external estimates of discards assuming 100% discard mortality, setting the maximum annual consumption of all prey by grey seals to (c) 1.5 mt and (d) 2.5 mt, and scaling the encounter rate by a factor of I 0.01 and (f) 100. Fishing mortality (F) includes all sources of fishing mortality (i.e., commercial fishing, bycatch from the groundfish fleet, and bycatch from the lobster fleet).*

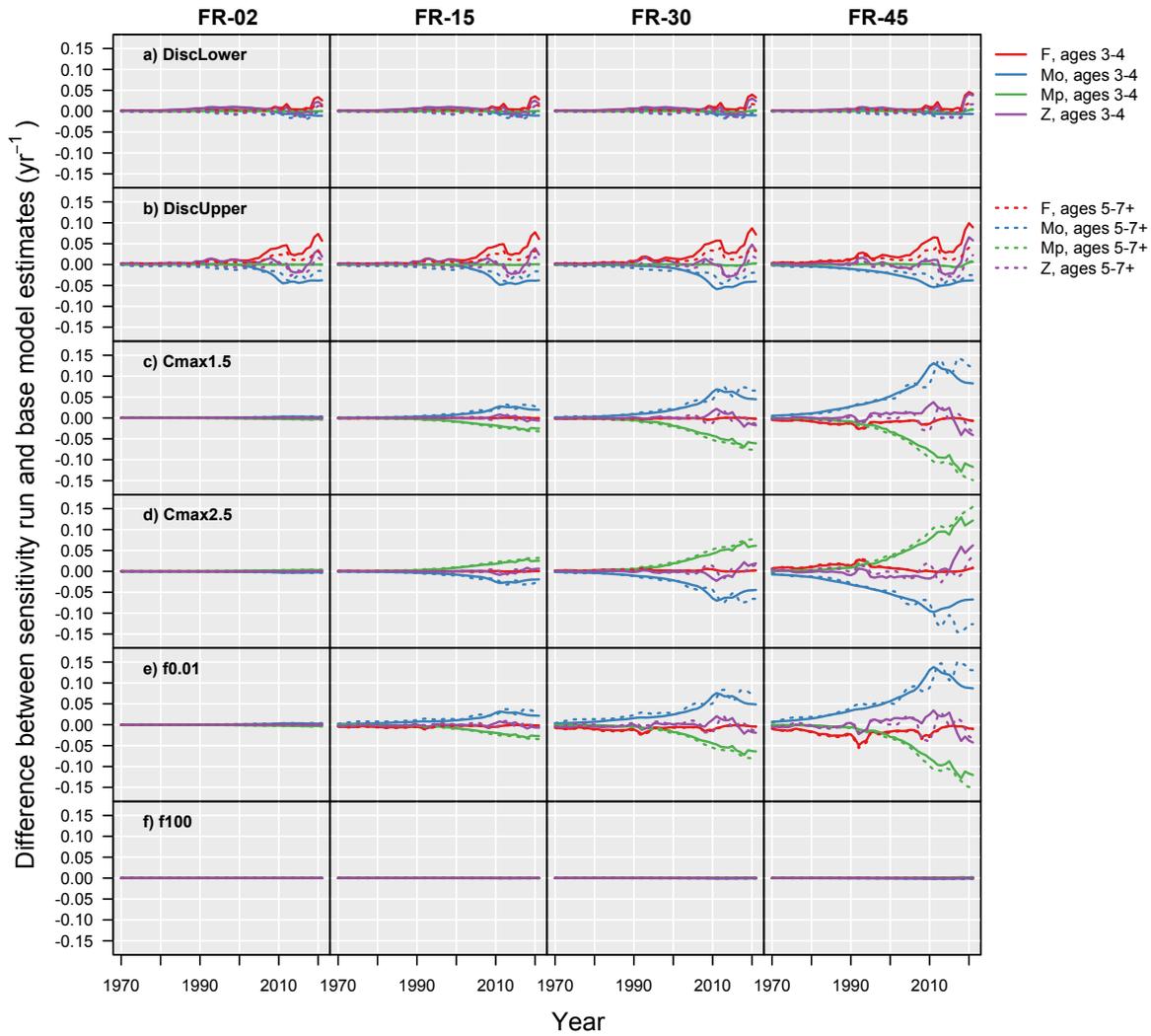

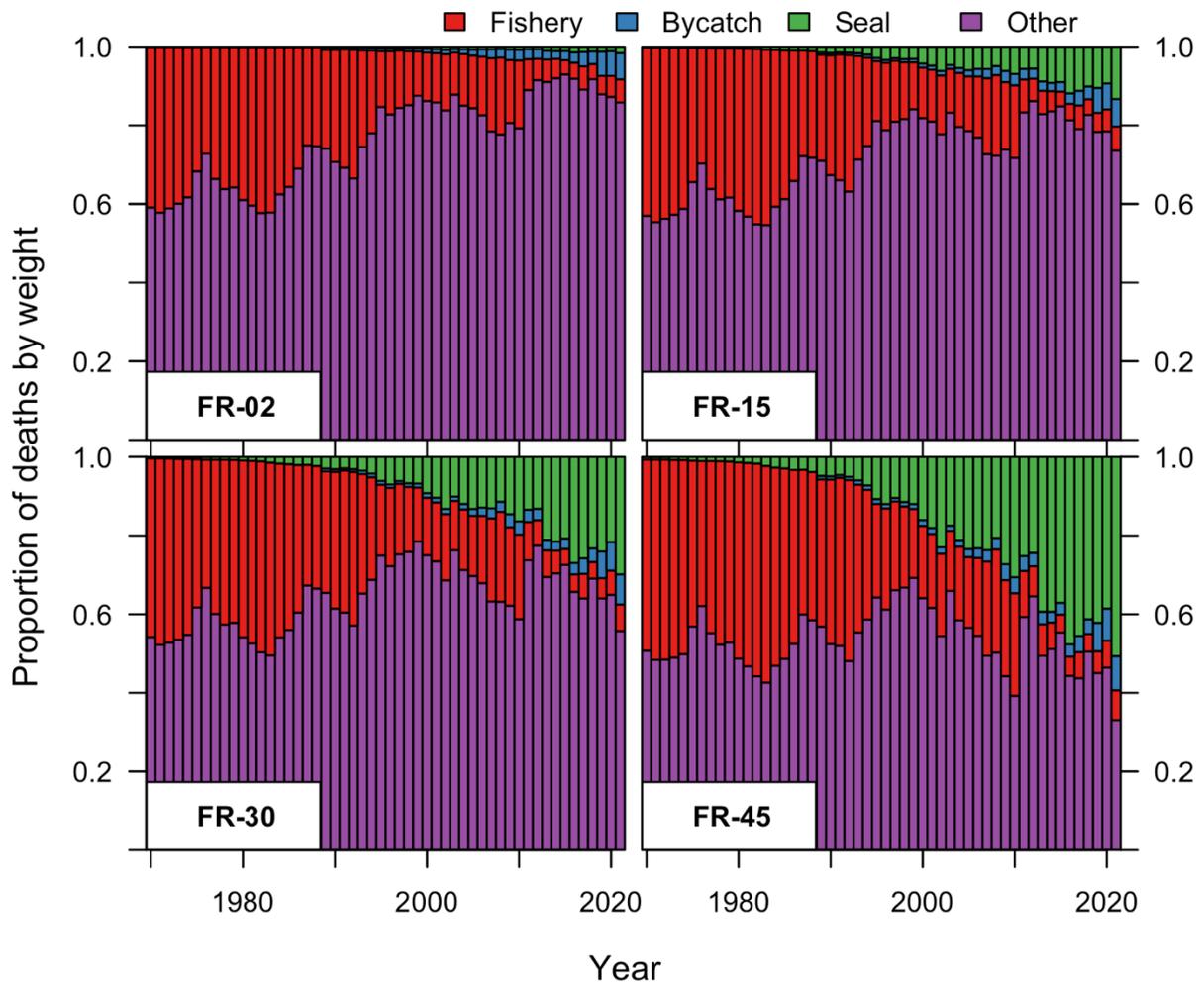

*Figure 9. Proportion of cod deaths in 4X5Y attributable to either directed, reported fishing (red), bycatch (blue), grey seal predation (green), and other causes (purple).*

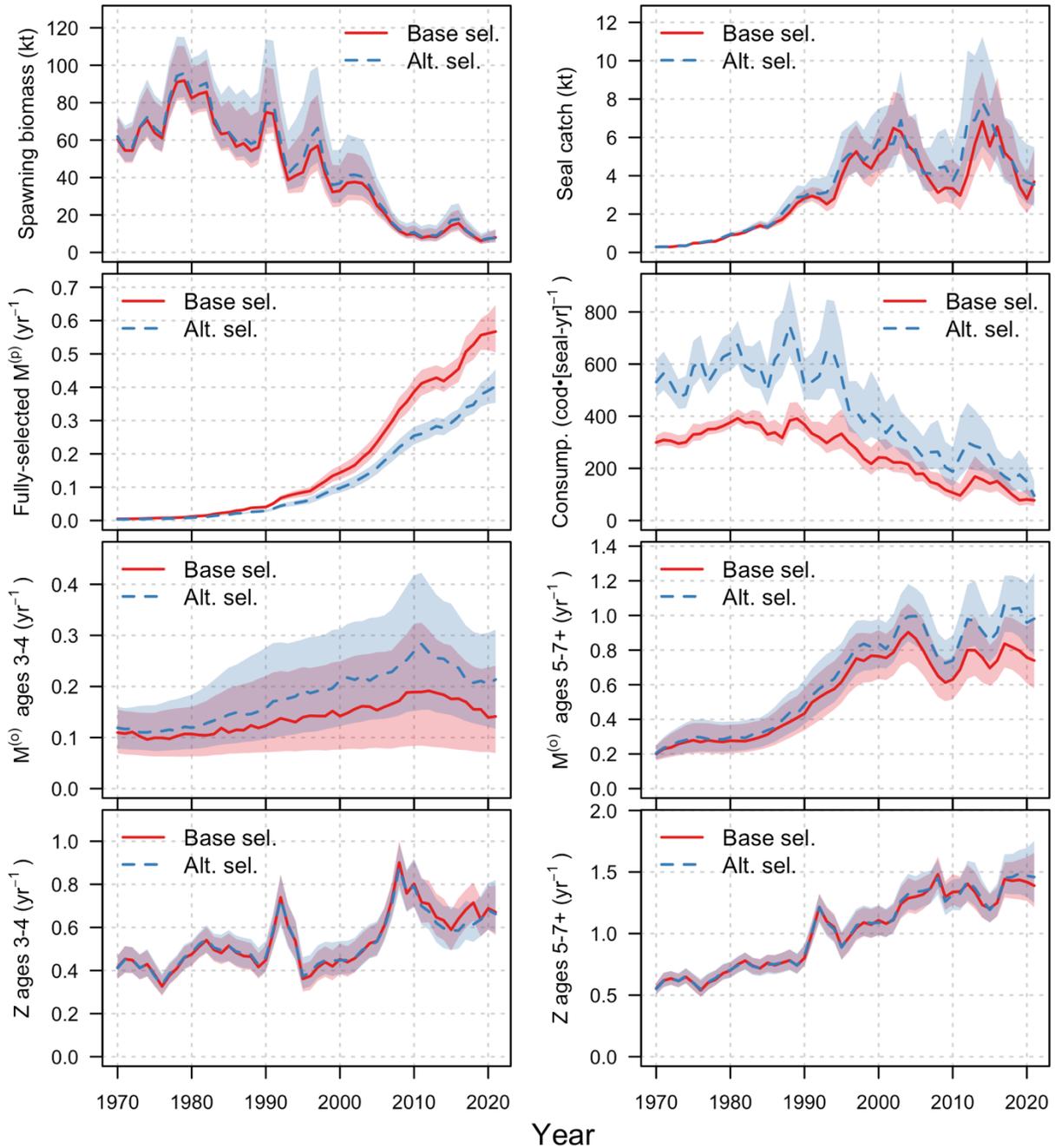

*Figure 10. Comparison of estimates of key population processes from FR-45 assuming that either the age-selectivity of cod to grey seal predation was equal to the proportion of cod mature-at-age (blue lines and shading) or assuming all cod age 2 and older were fully-selected by grey seals with recruit selectivity equal to 0.5 (red lines and shading). Lines are posterior modes while shaded regions are central 95% uncertainty intervals.*

# Appendix A

# Supplementary materials for Investigating the role of grey seal predation on 4X5Y cod productivity

**Tables**

Table A1. Proportion-at-age of Atlantic cod discards from 4X5Y Lobster (Homarus americanus) fisheries and associated sample size (N).

| Year | Age-1 | Age-2 | Age-3 | Age-4 | Age-5 | Age-6 | Age-7 | N |
|---|---|---|---|---|---|---|---|---|
| 2005 | 0.008 | 0.344 | 0.116 | 0.512 | 0.016 | 0.004 | 0.000 | 250 |
| 2006 | 0.031 | 0.290 | 0.515 | 0.083 | 0.049 | 0.031 | 0.000 | 324 |
| 2007 | 0.014 | 0.340 | 0.277 | 0.277 | 0.021 | 0.067 | 0.004 | 282 |
| 2008 | 0.010 | 0.249 | 0.290 | 0.133 | 0.195 | 0.123 | 0.000 | 293 |
| 2009 | 0.031 | 0.290 | 0.515 | 0.083 | 0.049 | 0.031 | 0.000 | 324 |
| 2010 | 0.023 | 0.250 | 0.320 | 0.377 | 0.010 | 0.007 | 0.013 | 300 |
| 2011 | 0.021 | 0.356 | 0.266 | 0.284 | 0.062 | 0.007 | 0.000 | 289 |
| 2012 | 0.030 | 0.405 | 0.320 | 0.201 | 0.041 | 0.003 | 0.000 | 338 |
| 2013 | 0.010 | 0.110 | 0.300 | 0.560 | 0.020 | 0.000 | 0.000 | 300 |
| 2014 | 0.004 | 0.105 | 0.371 | 0.511 | 0.009 | 0.000 | 0.000 | 229 |
| 2015 | 0.007 | 0.401 | 0.204 | 0.303 | 0.073 | 0.011 | 0.000 | 274 |
| 2016 | 0.020 | 0.242 | 0.657 | 0.051 | 0.024 | 0.003 | 0.003 | 297 |
| 2017 | 0.000 | 0.134 | 0.278 | 0.354 | 0.209 | 0.004 | 0.011 | 277 |
| 2018 | 0.018 | 0.152 | 0.287 | 0.366 | 0.146 | 0.006 | 0.024 | 164 |
| 2019 | 0.021 | 0.139 | 0.191 | 0.524 | 0.036 | 0.067 | 0.021 | 330 |
| 2020 | 0.021 | 0.144 | 0.354 | 0.144 | 0.337 | 0.000 | 0.000 | 291 |
| 2021 | 0.045 | 0.340 | 0.461 | 0.103 | 0.040 | 0.003 | 0.000 | 397 |

*Table A2. Average proportion of each month spent in NAFO area 4X by 208 satellite-tracked grey seals deployed between 1993 and 2014.*

| Month | Shelf | | Gulf | |
|---|---|---|---|---|
| | Male | Female | Male | Female |
| Jan | 0.1001 | 0.0037 | 0.0285 | 0.0000 |
| Feb | 0.3202 | 0.0000 | 0.0061 | 0.0000 |
| Mar | 0.3183 | 0.0000 | 0.0431 | 0.0000 |
| Apr | 0.1256 | 0.0000 | 0.0284 | 0.0000 |
| May | 0.0991 | 0.0000 | 0.0000 | 0.0000 |
| June | 0.0000 | 0.0000 | 0.0000 | 0.0000 |
| July | 0.0091 | 0.0000 | 0.0000 | 0.0000 |
| Aug | 0.0181 | 0.0000 | 0.0000 | 0.0000 |
| Sep | 0.0043 | 0.0000 | 0.0000 | 0.0000 |
| Oct | 0.0111 | 0.0000 | 0.0000 | 0.0000 |
| Nov | 0.0797 | 0.0004 | 0.0000 | 0.0000 |
| Dec | 0.1250 | 0.0019 | 0.0041 | 0.0000 |

*Table A3. Summary of literature on percent cod in grey seal diet (updated from O'Boyle & Sinclair 2012). N is the number of samples. Est. type describes whether the % cod estimate represents the contribution of cod to the grey seal diet by weight or by frequency.. Full citations are listed below the table. Abbreviations: GSL – Gulf of St. Lawrence*

| Location/popn. | Years | Method | % cod | Est. type | N | Reference |
|---|---|---|---|---|---|---|
| Iceland | 1979-1982 | Stomach | 22.0 | Weight | 97 | Hauksson (1985) |
| Iceland | 1992-1993 | Stomach | 25.1 | Weight | 737 | Hauksson and Bogason (1997) |
| British Seas East Coast | 1983-1988 | Scat | 21.6 | Weight | 236 | Hammond and Grellier (2006) |
| British Seas Orkney | 1985 | Scat | 5.1 | Weight | 859 | Hammond and Grellier (2006) |
| British Seas Donna Nook | 1985 | Scat | 12.1 | Weight | 360 | Hammond and Grellier (2006) |
| British Seas East Coast | 2002 | Scat | 8.2 | Weight | 429 | Hammond and Grellier (2006) |
| British Seas Orkney | 2002 | Scat | 10.2 | Weight | 711 | Hammond and Grellier (2006) |
| British Seas Shetland | 2002 | Scat | 7.7 | Weight | 244 | Hammond and Grellier (2006) |
| British Seas Donna Nook | 2002 | Scat | 4.5 | Weight | 429 | Hammond and Grellier (2006) |
| British Seas Inner Hebrides | 2010-2011 | Scat | 11.2 | Weight | 333 | Hammond and Wilson (2016) |
| British Seas Outer Hebrides | 2010-2011 | Scat | 10.7 | Weight | 274 | Hammond and Wilson (2016) |
| British Seas Shetland | 2010-2011 | Scat | 8.8 | Weight | 269 | Hammond and Wilson (2016) |
| British Seas Orkney | 2010-2011 | Scat | 8.9 | Weight | 742 | Hammond and Wilson (2016) |
| Central North Sea | 2010-2011 | Scat | 3.0 | Weight | 383 | Hammond and Wilson (2016) |
| Southern North Sea | 2000-2006 | Fatty acid | 0.3 | Weight | 15 | Gilles et al. (2008) |
| Southern North Sea | 2010-2011 | Scat | 3.1 | Weight | 204 | Hammond and Wilson (2016) |

| Location | Years | Sample | % | Measure | N | Reference |
|---|---|---|---|---|---|---|
| Southern North Sea | 2015-2017 | Scat | 0.1 | Freq | 21 | Boyi et al. (2022) |
| Baltic Sea Sweden | 2001-2005 | Gastroin. | 0.3 | Weight | 299 | Lundström et al. (2010) |
| Baltic Sea Poland | 2013 | Scat | 31 | Freq | 49 | Keszka et al. (2020) |
| Norwegian Sea | 1999-2010 | Scat & Gastroin. | 25.3 | Weight | 293 | Nilssen et al. (2019) |
| GSL upper | 1950-1987 | Stomach | 22.8 | Freq | 316 | Benoit and Bowen (1990) |
| GSL lower | 1950-1987 | Stomach | 13.5 | Freq | 89 | Benoit and Bowen (1990) |
| GSL Newfoundland inshore | 1985-2004 | Stomach | 4.2 | Weight | 25 | Hammill et al. (2007) |
| GSL Anticosti Island | 1988-1992 | Stomach | 17.8 | Weight | 183 | Hammill et al. (2007) |
| GSL lower | 1994-2003 | Stomach | 12.8 | Weight | 322 | Hammill et al. (2007) |
| Scotian Shelf coastal | 1950-1987 | Stomach | 13.6 | Freq | 213 | Benoit and Bowen (1990) |
| Scotian Shelf Sable Island | 1950-1987 | Stomach | 21.3 | Weight | 47 | Benoit and Bowen (1990) |
| Scotian Shelf inshore | 1988-1990 | Stomach | 17 | Weight | 106 | Bowen et al. (1993) |
| Scotian Shelf Sable Island | 1988-1990 | Stomach | 10.3 | Weight | 37 | Bowen et al. (1993) |
| Scotian Shelf Sable Island | 1991-1998 | Scat | 7 | Weight | 1304 | Bowen and Harrison (2007) and Bowen et al. (2011) |
| Scotian Shelf Sable Island | 1993-2001 | Fatty acid | 1.9 | Weight | 587 | Beck et al. (2011) |
| Gulf of Maine | 2004-2008 | Scat | 6.4 | Weight | 305 | Ampela (2009) |
| Gulf of Maine | 1998-2004 | Stomach | 1.7 | Weight | 49 | Ampela (2009) |
| GSL Cape Breton Island (Males) | 1996-2011 | Stomach | 11.1 | Weight | 54 | Hammill et al. (2014) |

| Location | Years | Sample | Value | Metric | N | Reference |
|---|---|---|---|---|---|---|
| GSL Cape Breton Island (Males) | 1996-2011 | Intestine | 10.8 | Weight | 98 | Hammill et al. (2014) |
| GSL Cape Breton Island (Females) | 1996-2011 | Stomach | 26.5 | Weight | 40 | Hammill et al. (2014) |
| GSL Cape Breton Island (Females) | 1996-2011 | Intestine | 6.4 | Weight | 71 | Hammill et al. (2014) |
| GSL Cabot Strait (Males) | 2008-2011 | Stomach | 68.4 | Weight | 81 | Hammill et al. (2014) |
| GSL Cabot Strait (Males) | 2008-2011 | Intestine | 46.5 | Weight | 166 | Hammill et al. (2014) |
| GSL Cabot Strait (Females) | 2008-2011 | Stomach | 14.3 | Weight | 25 | Hammill et al. (2014) |
| GSL Cabot Strait (Females) | 2008-2011 | Intestine | 9.4 | Weight | 37 | Hammill et al. (2014) |

Table A4. Results from beta regression of the observed proportion of cod in grey seal diets on study factors, including the region (east or west Atlantic), sample size (N), sampling method, study year (set to the midpoint between start year and end year for multi-year studies), and observation type (by weight or frequency). The base levels for region, method and type were east, fatty acid analysis, and frequency, respectively. Pseudo $R^2=0.380$.

|  | Estimate | Std. Error | z value | Pr(>\|z\|) |
|---|---|---|---|---|
| (Intercept) | 6.059 | 24.009 | 0.252 | 0.801 |
| Region [West] | -0.280 | 0.389 | -0.721 | 0.471 |
| N | 0.000 | 0.001 | 0.756 | 0.450 |
| Method [Gastoint.] | -0.792 | 1.406 | -0.563 | 0.573 |
| Method [Intestine] | 1.579 | 1.025 | 1.539 | 0.124 |
| Method [Scat] | 0.599 | 1.010 | 0.593 | 0.553 |
| Method [Scat & Gastroint.] | 1.796 | 1.201 | 1.496 | 0.135 |
| Method [Stomach] | 1.566 | 0.979 | 1.600 | 0.110 |
| Year | -0.005 | 0.012 | -0.380 | 0.704 |
| TypeWeight | 0.230 | 0.422 | 0.545 | 0.586 |

*Table A5. MCMC diagnostics for each model including the maximum potential scale reduction factor on rank- normalized split chains for all parameters (), the minimum effective sample size of the rank-normalized draws for all parameters (BulkESS), the minimum of the effective sample sizes of the 5% and 95% quantiles for all parameters (TailESS), the number of divergent transitions (#DT), the number of transitions that saturated the maximum tree depth (#SMT), and the minimum energy Bayesian fraction of missing information for all chains (E-BFMI).  values exceeding 1.01 indicate problems with convergence. Effective sample sizes of at least 300 (100 per chain) were considered acceptable. E-BFMI values below 0.3 indicate that HMC may have had difficulty exploring the target distribution.*

| Model |       | BulkESS | TailESS | #DT | #SMT | E-BFMI |
|-------|-------|---------|---------|-----|------|--------|
| RW    | 1.007 | 1205    | 1358    | 0   | 0    | 0.963  |
| FR-02 | 1.001 | 1588    | 1999    | 0   | 0    | 0.907  |
| FR-15 | 1.001 | 1683    | 2632    | 3   | 0    | 1.001  |
| FR-30 | 1.002 | 1561    | 1861    | 0   | 0    | 0.967  |
| FR-45 | 1.002 | 1677    | 1723    | 1   | 0    | 0.995  |

Table A6. Mean estimated predation mortality for 2017-2021 for five models (columns) and thirteen model scenarios (rows) including the base model and twelve alternative model which were run to test the sensitivity of base model estimates to particular assumptions. Alternative scenarios were defined by setting discards to the lower and upper bound of external estimates of discards with discard mortality (DM) set to either 100% or 25%, varying the maximum annual consumption of all prey by grey seals (; kt), and scaling the encounter rate (f). Predation mortality was averaged over 3-4 yr and 5-7 yr age-blocks.

| Scenario | FR-02 | | FR-15 | | FR-30 | | FR-45 | |
|---|---|---|---|---|---|---|---|---|
| | $a$=3:4 | $a$=5:7 | $a$=3:4 | $a$=5:7 | $a$=3:4 | $a$=5:7 | $a$=3:4 | $a$=5:7 |
| Base | 0.01 | 0.02 | 0.11 | 0.13 | 0.26 | 0.31 | 0.48 | 0.57 |
| DiscLow, DM=100% | 0.01 | 0.02 | 0.11 | 0.13 | 0.26 | 0.31 | 0.48 | 0.58 |
| DiscLow, DM=25% | 0.01 | 0.01 | 0.11 | 0.13 | 0.25 | 0.30 | 0.48 | 0.57 |
| DiscUpp, DM=100% | 0.01 | 0.02 | 0.11 | 0.13 | 0.26 | 0.31 | 0.49 | 0.58 |
| DiscUpp, DM=25% | 0.01 | 0.01 | 0.11 | 0.13 | 0.25 | 0.30 | 0.48 | 0.57 |
| $C_{max}$ = 1.50 | 0.01 | 0.01 | 0.08 | 0.09 | 0.18 | 0.22 | 0.34 | 0.41 |
| $C_{max}$ = 1.75 | 0.01 | 0.01 | 0.09 | 0.11 | 0.22 | 0.26 | 0.40 | 0.48 |
| $C_{max}$ = 2.25 | 0.01 | 0.02 | 0.12 | 0.14 | 0.28 | 0.33 | 0.52 | 0.62 |
| $C_{max}$ = 2.50 | 0.02 | 0.02 | 0.13 | 0.16 | 0.31 | 0.37 | 0.58 | 0.69 |
| $f = f^*/100$ | 0.01 | 0.01 | 0.08 | 0.09 | 0.19 | 0.22 | 0.35 | 0.42 |
| $f = f^*/10$ | 0.01 | 0.01 | 0.10 | 0.12 | 0.25 | 0.30 | 0.46 | 0.55 |
| $f = f^* \times 10$ | 0.01 | 0.02 | 0.11 | 0.13 | 0.26 | 0.31 | 0.48 | 0.57 |
| $f = f^* \times 100$ | 0.01 | 0.02 | 0.11 | 0.13 | 0.26 | 0.31 | 0.48 | 0.58 |

Table A7. Mean estimated "other" (i.e., non-predation) natural mortality for 2017-2021 for five models (columns) and thirteen model scenarios (rows) including the base model and twelve alternative model which were run to test the sensitivity of base model estimates to particular assumptions. Alternative scenarios were defined by setting discards to the lower and upper bound of external estimates of discards with discard mortality (DM) set to either 100% or 25%, varying the maximum annual consumption of all prey by grey seals (; kt), and scaling the encounter rate (f). Predation mortality was averaged over 3-4 yr and 5-7 yr age-blocks.

| Scenario | FR-02 | | FR-15 | | FR-30 | | FR-45 | |
|---|---|---|---|---|---|---|---|---|
| | $a$=3:4 | $a$=5:7 | $a$=3:4 | $a$=5:7 | $a$=3:4 | $a$=5:7 | $a$=3:4 | $a$=5:7 |
| Base | 0.01 | 0.02 | 0.11 | 0.13 | 0.26 | 0.31 | 0.48 | 0.57 |
| DiscLow, DM=100% | 0.01 | 0.02 | 0.11 | 0.13 | 0.26 | 0.31 | 0.48 | 0.58 |
| DiscLow, DM=25% | 0.01 | 0.01 | 0.11 | 0.13 | 0.25 | 0.30 | 0.48 | 0.57 |
| DiscUpp, DM=100% | 0.01 | 0.02 | 0.11 | 0.13 | 0.26 | 0.31 | 0.49 | 0.58 |
| DiscUpp, DM=25% | 0.01 | 0.01 | 0.11 | 0.13 | 0.25 | 0.30 | 0.48 | 0.57 |
| $C_{max} = 1.50$ | 0.51 | 1.34 | 0.46 | 1.25 | 0.37 | 1.12 | 0.25 | 0.94 |
| $C_{max} = 1.75$ | 0.51 | 1.33 | 0.45 | 1.24 | 0.35 | 1.09 | 0.20 | 0.87 |
| $C_{max} = 2.25$ | 0.51 | 1.33 | 0.43 | 1.21 | 0.30 | 1.02 | 0.12 | 0.74 |
| $C_{max} = 2.50$ | 0.51 | 1.33 | 0.42 | 1.19 | 0.28 | 0.98 | 0.09 | 0.67 |
| $f = f^*/100$ | 0.51 | 1.34 | 0.46 | 1.26 | 0.37 | 1.13 | 0.25 | 0.95 |
| $f = f^*/10$ | 0.51 | 1.33 | 0.44 | 1.23 | 0.33 | 1.06 | 0.17 | 0.82 |
| $f = f^* \times 10$ | 0.51 | 1.33 | 0.44 | 1.22 | 0.32 | 1.05 | 0.16 | 0.80 |
| $f = f^* \times 100$ | 0.51 | 1.33 | 0.44 | 1.22 | 0.32 | 1.05 | 0.16 | 0.80 |

*Table A8. Mean estimated fishing mortality for 2017-2021 for five models (columns) and thirteen model scenarios (rows) including the base model and twelve alternative model which were run to test the sensitivity of base model estimates to particular assumptions. Alternative scenarios were defined by setting discards to the lower and upper bound of external estimates of discards with discard mortality (DM) set to either 100% or 25%, varying the maximum annual consumption of all prey by grey seals (; kt), and scaling the encounter rate (f). Predation mortality was averaged over 3-4 yr and 5-7 yr age-blocks.*

| Scenario | FR-02 | | FR-15 | | FR-30 | | FR-45 | |
|---|---|---|---|---|---|---|---|---|
| | $a$=3:4 | $a$=5:7 | $a$=3:4 | $a$=5:7 | $a$=3:4 | $a$=5:7 | $a$=3:4 | $a$=5:7 |
| Base | 0.054 | 0.062 | 0.057 | 0.066 | 0.064 | 0.074 | 0.074 | 0.086 |
| DiscLow, DM=100% | 0.073 | 0.070 | 0.078 | 0.074 | 0.088 | 0.084 | 0.099 | 0.095 |
| DiscLow, DM=25% | 0.058 | 0.064 | 0.062 | 0.068 | 0.069 | 0.077 | 0.080 | 0.088 |
| DiscUpp, DM=100% | 0.108 | 0.082 | 0.114 | 0.087 | 0.127 | 0.098 | 0.141 | 0.108 |
| DiscUpp, DM=25% | 0.067 | 0.067 | 0.071 | 0.072 | 0.080 | 0.081 | 0.091 | 0.092 |
| $C_{max} = 1.50$ | 0.060 | 0.070 | 0.050 | 0.060 | 0.060 | 0.070 | 0.090 | 0.100 |
| $C_{max} = 1.75$ | 0.060 | 0.070 | 0.050 | 0.060 | 0.060 | 0.070 | 0.090 | 0.110 |
| $C_{max} = 2.25$ | 0.060 | 0.070 | 0.050 | 0.060 | 0.060 | 0.070 | 0.100 | 0.110 |
| $C_{max} = 2.50$ | 0.060 | 0.070 | 0.050 | 0.060 | 0.060 | 0.070 | 0.100 | 0.120 |
| $f = f^* / 100$ | 0.059 | 0.067 | 0.048 | 0.056 | 0.057 | 0.066 | 0.086 | 0.099 |
| $f = f^* / 10$ | 0.059 | 0.067 | 0.049 | 0.057 | 0.060 | 0.069 | 0.094 | 0.109 |
| $f = f^* \times 10$ | 0.059 | 0.067 | 0.049 | 0.057 | 0.060 | 0.069 | 0.095 | 0.110 |
| $f = f^* \times 100$ | 0.059 | 0.067 | 0.049 | 0.057 | 0.060 | 0.069 | 0.096 | 0.110 |

Table A9. Mean estimated total mortality in 2017-2021 for five models (columns) and thirteen model scenarios (rows) including the base model and twelve alternative model which were run to test the sensitivity of base model estimates to particular assumptions. Alternative scenarios were defined by setting discards to the lower and upper bound of external estimates of discards with discard mortality (DM) set to either 100% or 25%, varying the maximum annual consumption of all prey by grey seals (; kt), and scaling the encounter rate (f). Predation mortality was averaged over 3-4 yr and 5-7 yr age-blocks.

| Scenario | FR-02 | | FR-15 | | FR-30 | | FR-45 | |
|---|---|---|---|---|---|---|---|---|
| | $a$=3:4 | $a$=5:7 | $a$=3:4 | $a$=5:7 | $a$=3:4 | $a$=5:7 | $a$=3:4 | $a$=5:7 |
| Base | 0.58 | 1.41 | 0.60 | 1.41 | 0.63 | 1.42 | 0.69 | 1.44 |
| DiscLow, DM=100% | 0.62 | 1.36 | 0.64 | 1.36 | 0.68 | 1.37 | 0.76 | 1.38 |
| DiscLow, DM=25% | 0.61 | 1.36 | 0.64 | 1.36 | 0.67 | 1.37 | 0.74 | 1.38 |
| DiscUpp, DM=100% | 0.63 | 1.36 | 0.65 | 1.36 | 0.69 | 1.37 | 0.77 | 1.38 |
| DiscUpp, DM=25% | 0.62 | 1.36 | 0.64 | 1.36 | 0.67 | 1.37 | 0.75 | 1.38 |
| $C_{max} = 1.50$ | 0.58 | 1.41 | 0.59 | 1.41 | 0.62 | 1.42 | 0.66 | 1.43 |
| $C_{max} = 1.75$ | 0.58 | 1.41 | 0.60 | 1.41 | 0.63 | 1.42 | 0.67 | 1.43 |
| $C_{max} = 2.25$ | 0.58 | 1.41 | 0.60 | 1.41 | 0.64 | 1.42 | 0.71 | 1.44 |
| $C_{max} = 2.50$ | 0.58 | 1.41 | 0.60 | 1.41 | 0.65 | 1.43 | 0.74 | 1.45 |
| $f = f^*/100$ | 0.58 | 1.41 | 0.59 | 1.41 | 0.62 | 1.42 | 0.66 | 1.43 |
| $f = f^*/10$ | 0.58 | 1.41 | 0.60 | 1.41 | 0.63 | 1.42 | 0.69 | 1.44 |
| $f = f^* \times 10$ | 0.58 | 1.41 | 0.60 | 1.41 | 0.63 | 1.42 | 0.69 | 1.44 |
| $f = f^* \times 100$ | 0.58 | 1.41 | 0.60 | 1.41 | 0.63 | 1.42 | 0.69 | 1.44 |

*Table A10. Mean estimated spawning biomass (kt) in 2017-2021 for five models (columns) and thirteen model scenarios (rows) including the base model and twelve alternative model which were run to test the sensitivity of base model estimates to particular assumptions. Alternative scenarios were defined by setting discards to the lower and upper bound of external estimates of discards with discard mortality (DM) set to either 100% or 25%, varying the maximum annual consumption of all prey by grey seals (; kt), and scaling the encounter rate (f). Predation mortality was averaged over 3-4 yr and 5-7 yr age-blocks.*

| Scenario | FR-02 | FR-15 | FR-30 | FR-45 |
|---|---|---|---|---|
| Base | 11.02 | 10.52 | 9.54 | 8.36 |
| DiscLow, DM=100% | 11.35 | 10.77 | 9.72 | 8.87 |
| DiscLow, DM=25% | 11.28 | 10.67 | 9.58 | 8.63 |
| DiscUpp, DM=100% | 11.39 | 10.85 | 9.85 | 9.20 |
| DiscUpp, DM=25% | 11.27 | 10.68 | 9.61 | 8.69 |
| $C_{max} = 1.50$ | 11.02 | 10.52 | 9.54 | 8.36 |
| $C_{max} = 1.75$ | 11.35 | 10.77 | 9.72 | 8.87 |
| $C_{max} = 2.25$ | 11.28 | 10.67 | 9.58 | 8.63 |
| $C_{max} = 2.50$ | 11.39 | 10.85 | 9.85 | 9.20 |
| $f = f^*/100$ | 11.27 | 10.68 | 9.61 | 8.69 |
| $f = f^*/10$ | 11.02 | 10.52 | 9.54 | 8.36 |
| $f = f^* \times 10$ | 11.35 | 10.77 | 9.72 | 8.87 |
| $f = f^* \times 100$ | 11.28 | 10.67 | 9.58 | 8.63 |

44 **Figures**

45 *Figure A1. Observed fishery landings of cod in 4X5Y (bars) and the range of estimated discards (lines).*

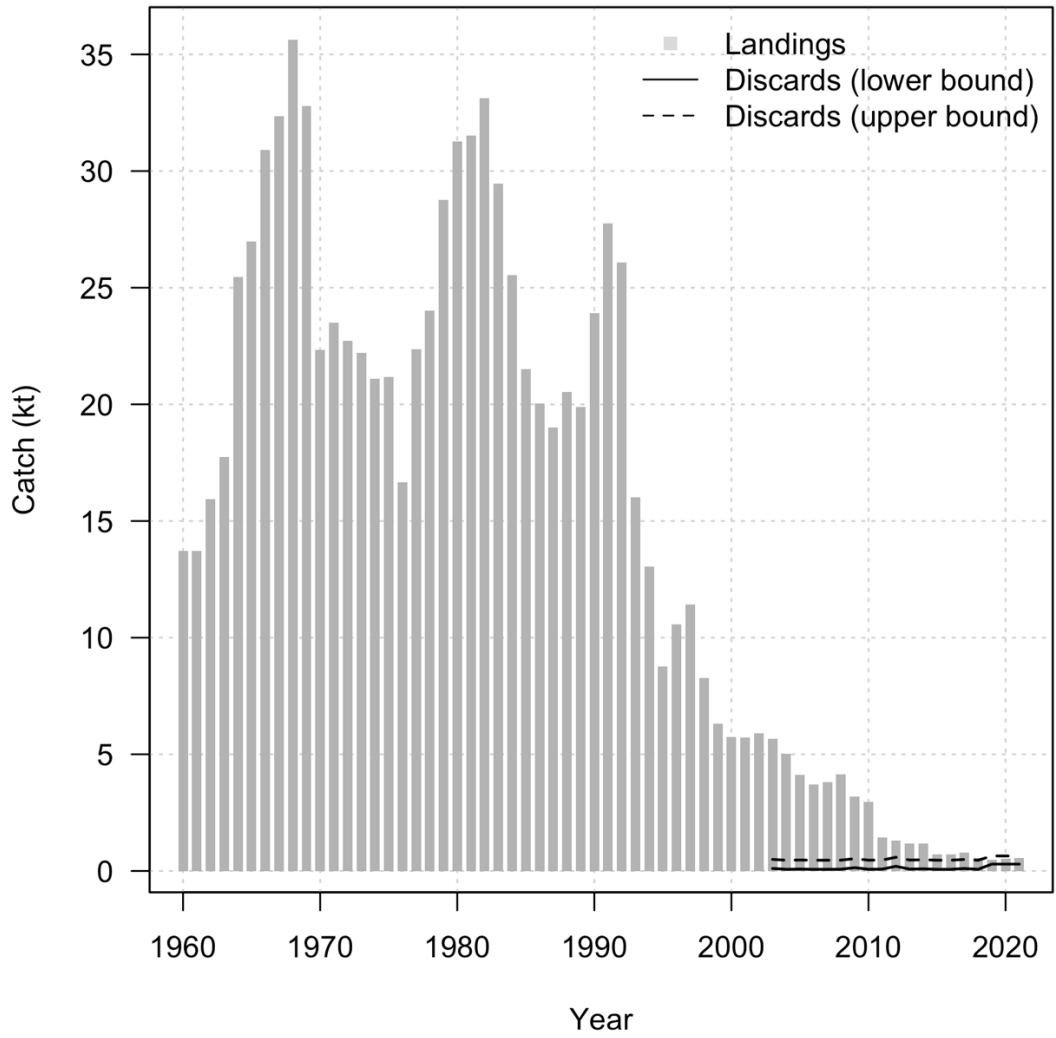

*Figure A2. Biomass indices for 4X5Y cod (Scotian Shelf and Bay of Fundy combined) from three research vessels, 1970-2021.*

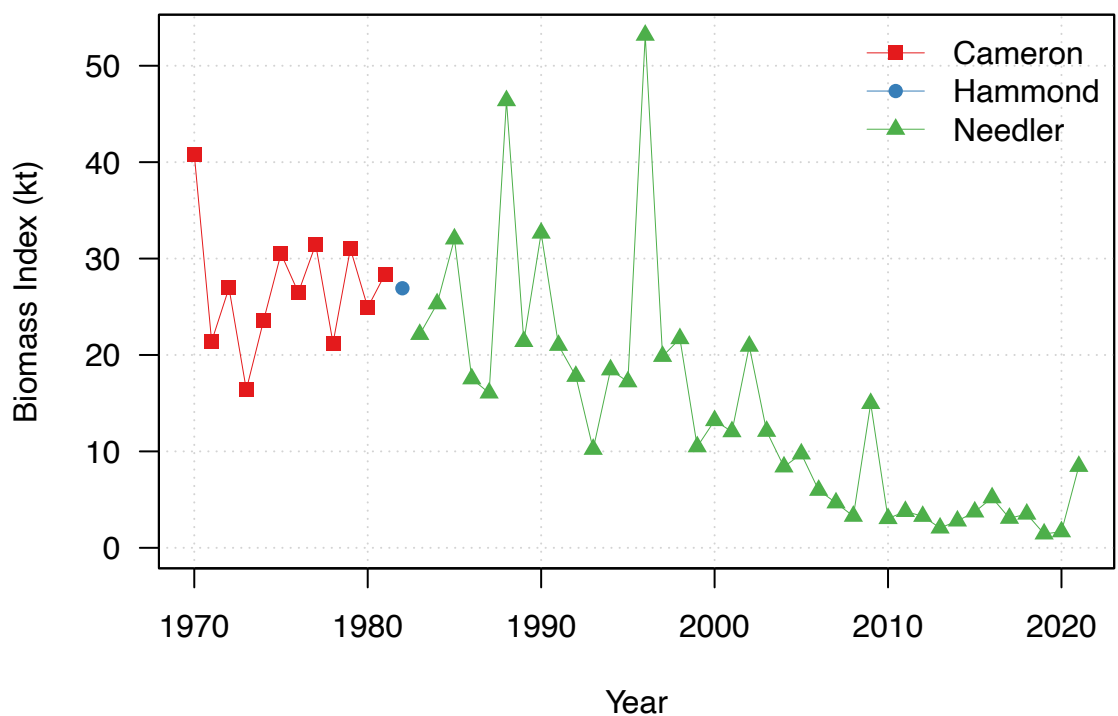

*Figure A3. Annual age-composition (proportion-at-age) of 4X5Y cod landed in the commercial fishery and RV survey, 1970-2021*

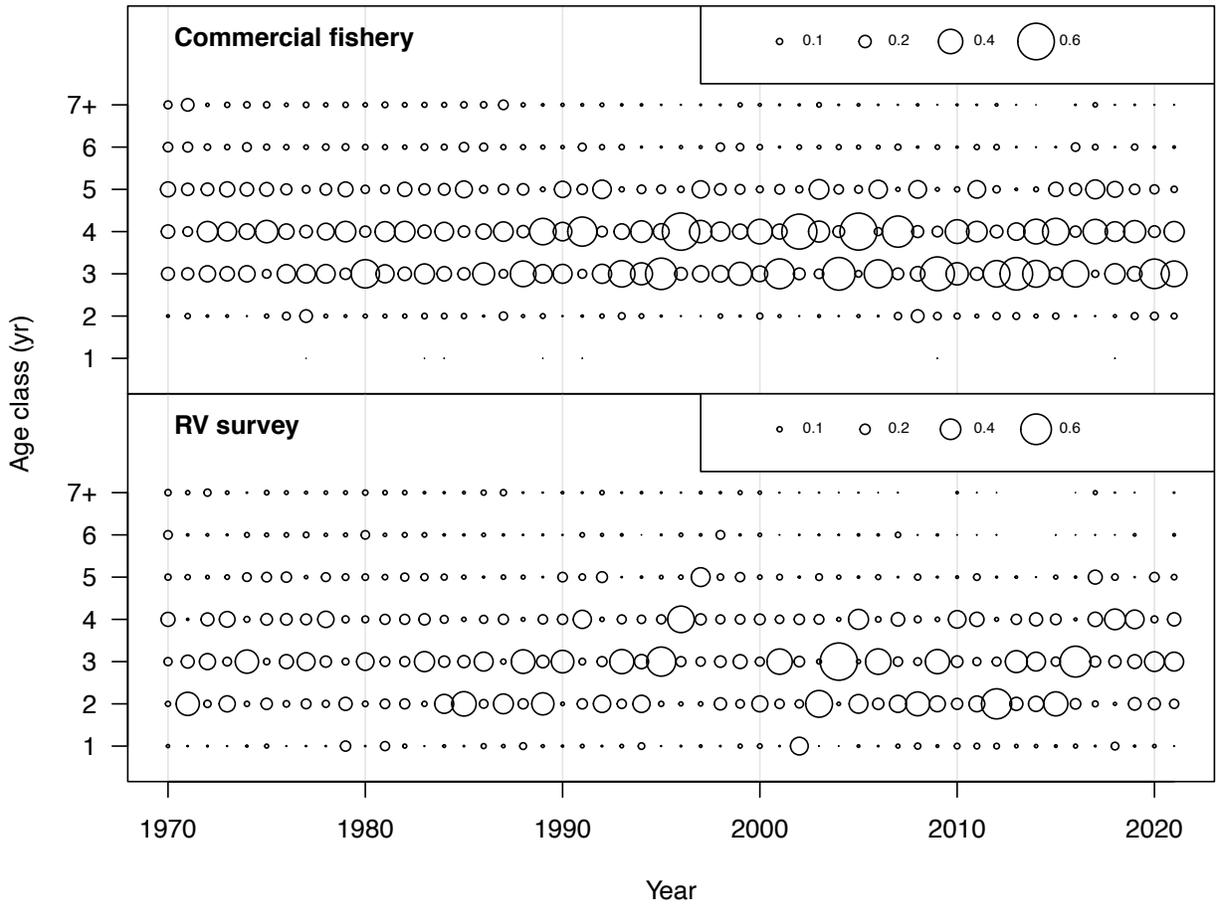

*Figure A4. Proportion of 4X5Y cod mature at each age, accounting for different maturity schedules on the Scotian Shelf and in the Bay of Fundy.*

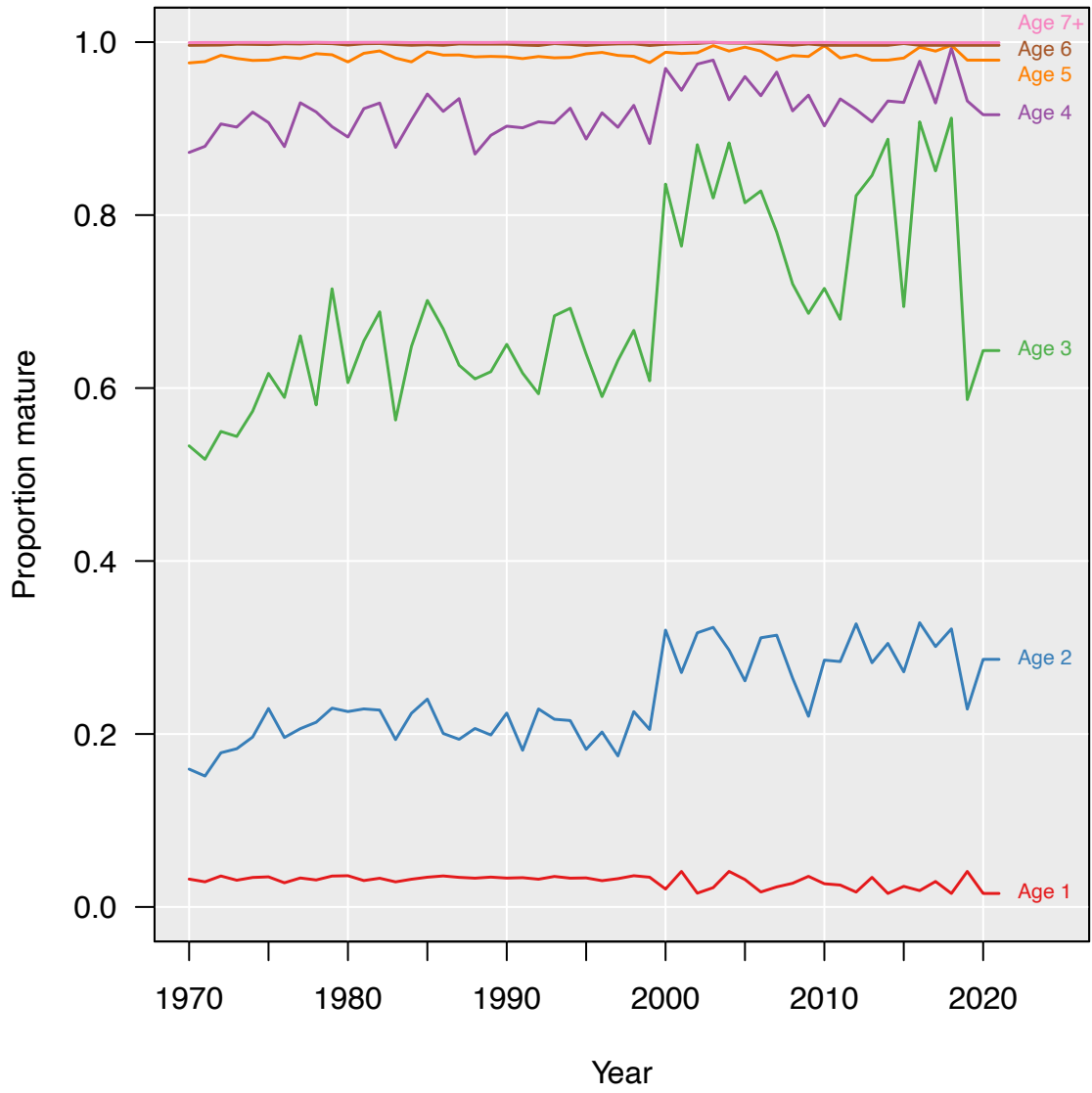

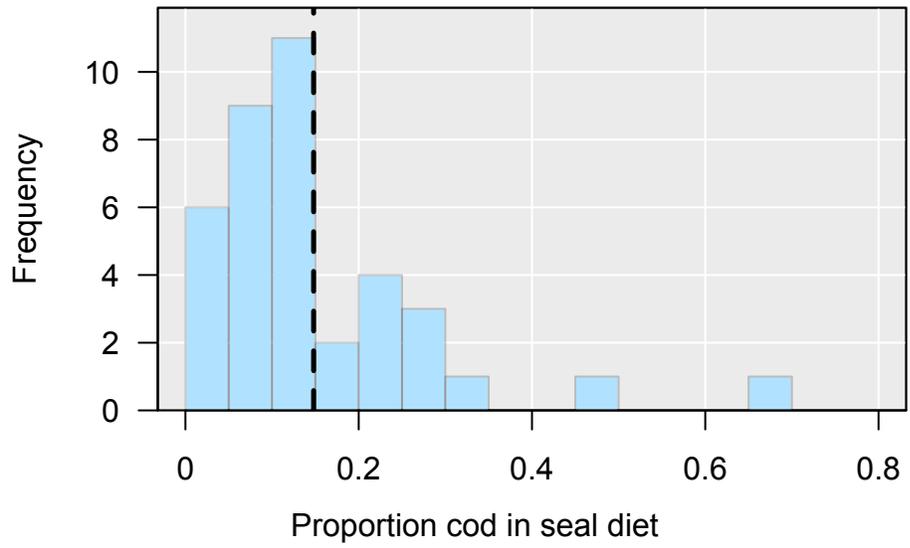

*Figure A5. Histogram of the proportional contribution of cod to the grey seal diet by weight from global diet studies. The dashed vertical line indicates the mean.*

*Figure A6. Model fits to RV survey biomass indices for 4X5Y cod (top panel) and grey seal foraging presence in 4X5Y (bottom panel). Circles represent input data while lines represent posterior modes.*

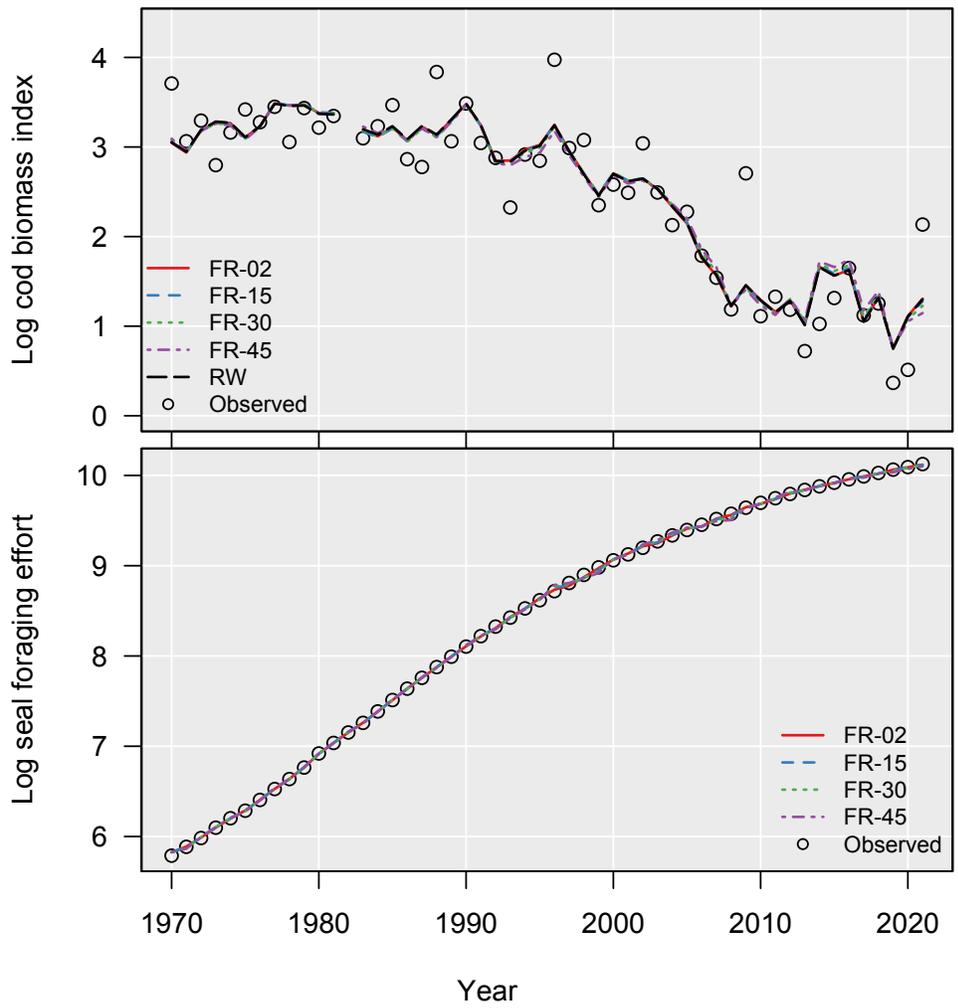

*Figure A7. Residuals (log(predicted) – log(observed)) for the commercial (top row) and RV survey (bottom row) age-composition from the RW model (left column) and FR-15 model (right column).*

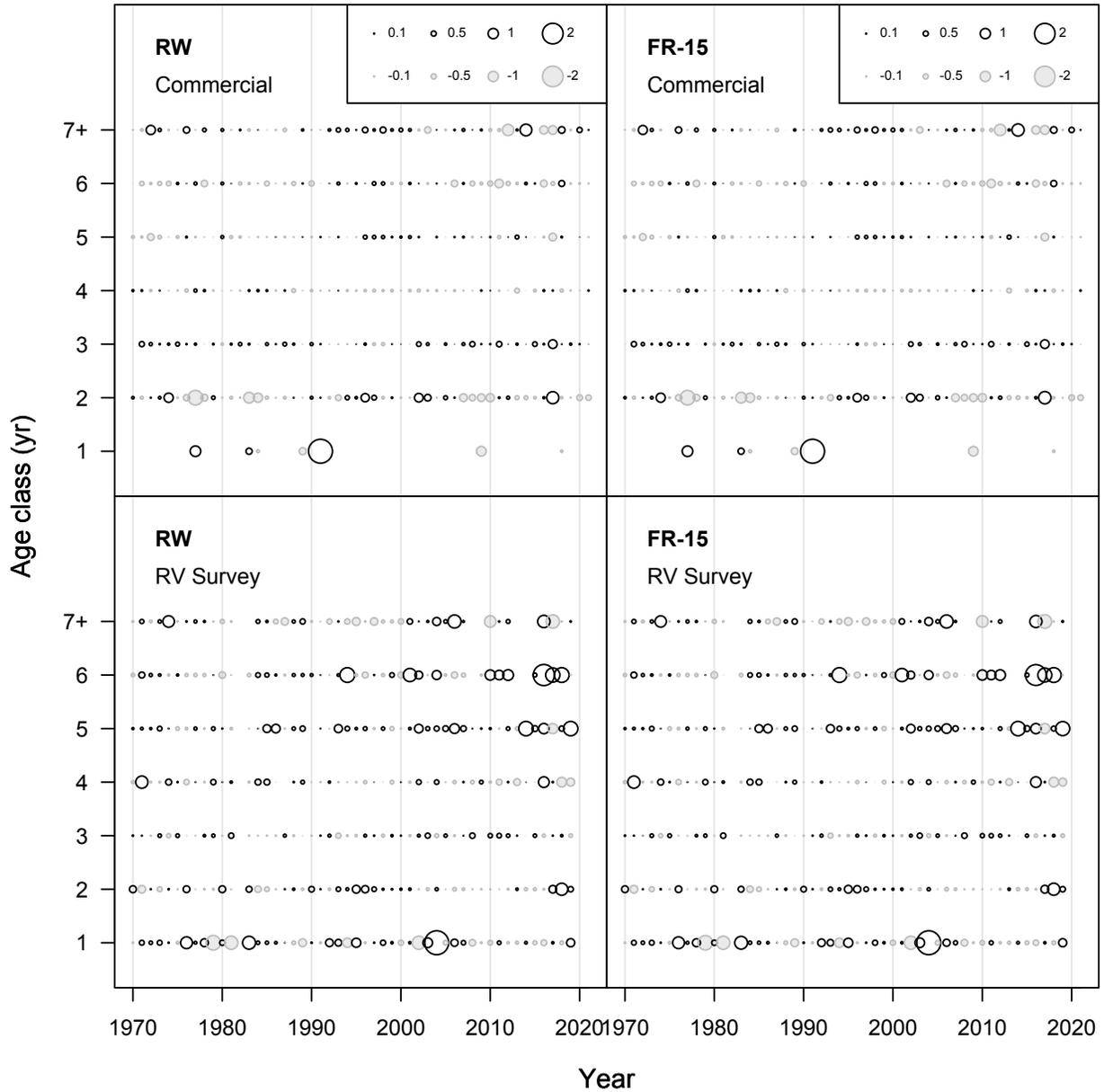

*Figure A8. Annual consumption of 4X5Y cod by grey seals assuming that the proportional contribution of cod to the grey seal diet by weight is either 0.02, 0.15, 0.30, or 0.45.*

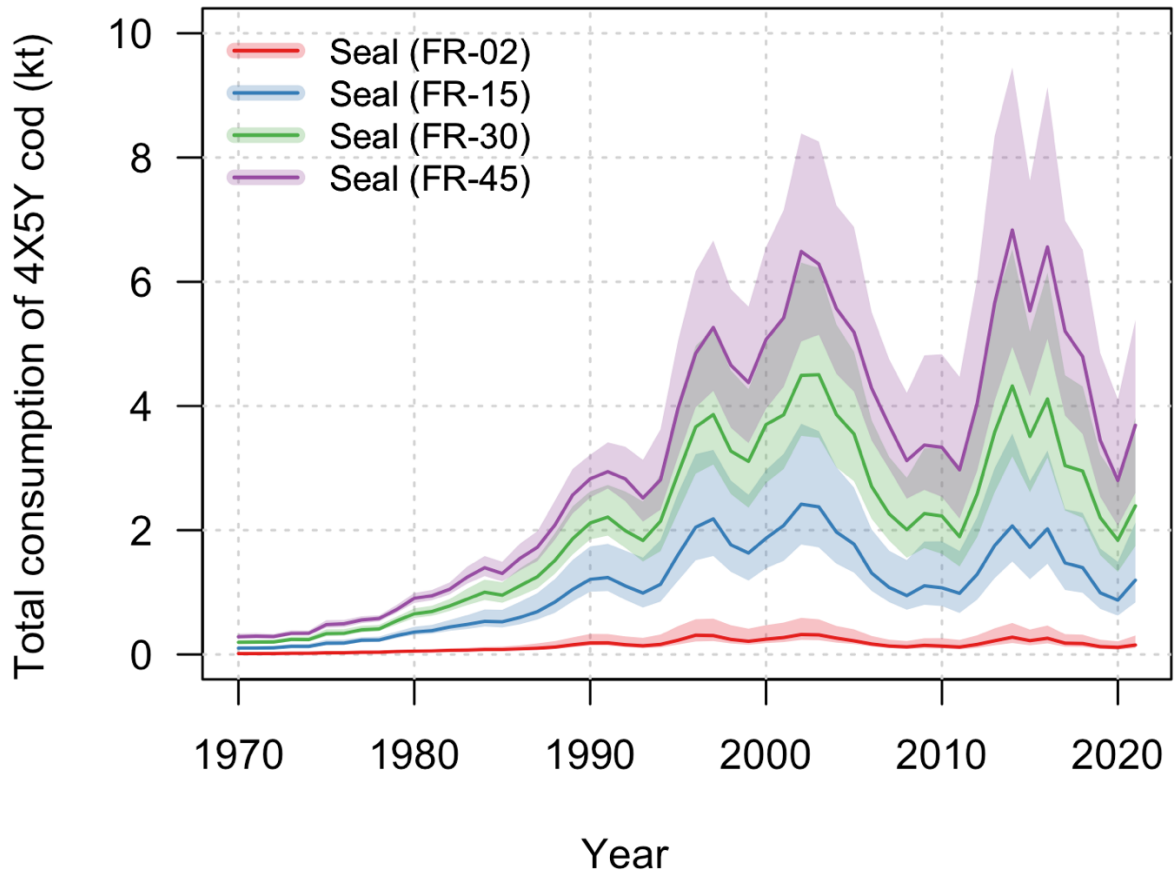

*Figure A9. Estimated annual removals/deaths of 4X5Y cod arising from fishing (F; including directed landings and bycatch), grey seal predation ($M_p$), and other natural mortality ($M_o$) from four models that were fit assuming either no discarding (Base) or that model discards were set to the upper bound of external discard estimates with 100% discarding mortality (DiscUpp).*

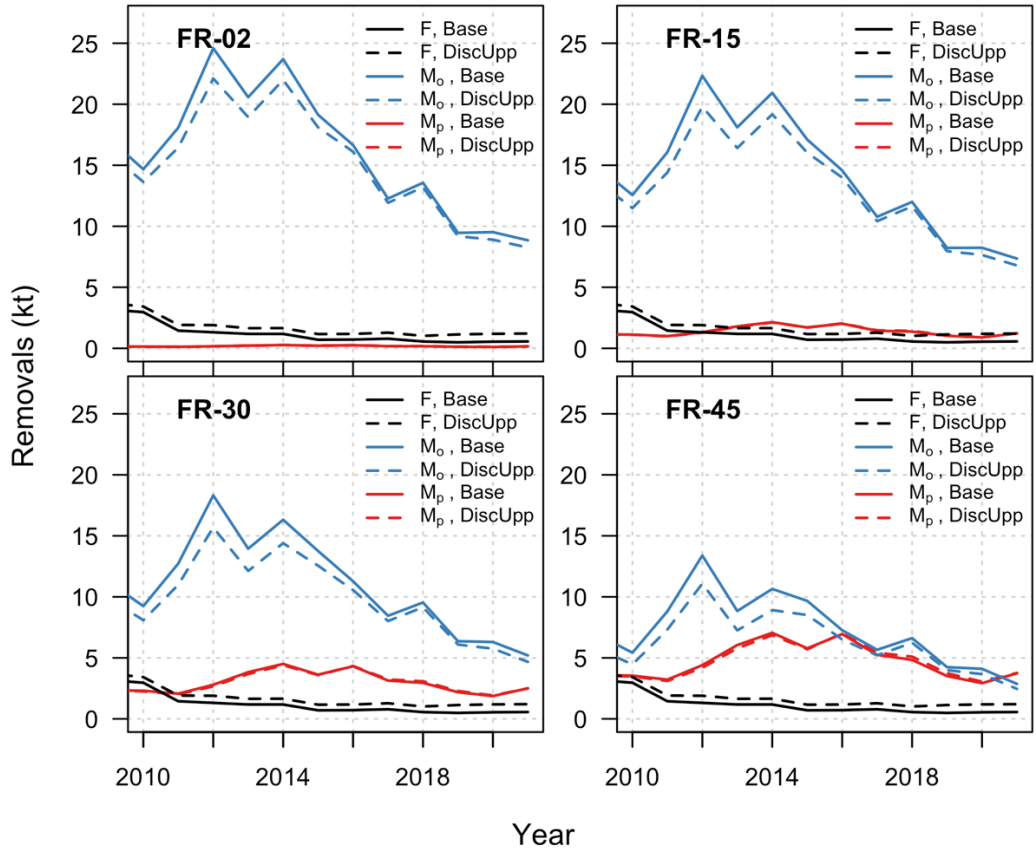

*Figure A10. Estimated spawning stock biomass (top row), recruitment (middle row), and recruits-per-spawner (bottom row) of 4X5Y cod from six statistical catch-at-age models (columns). Lines represent posterior modes while shaded regions represent the central 95% uncertainty interval.*

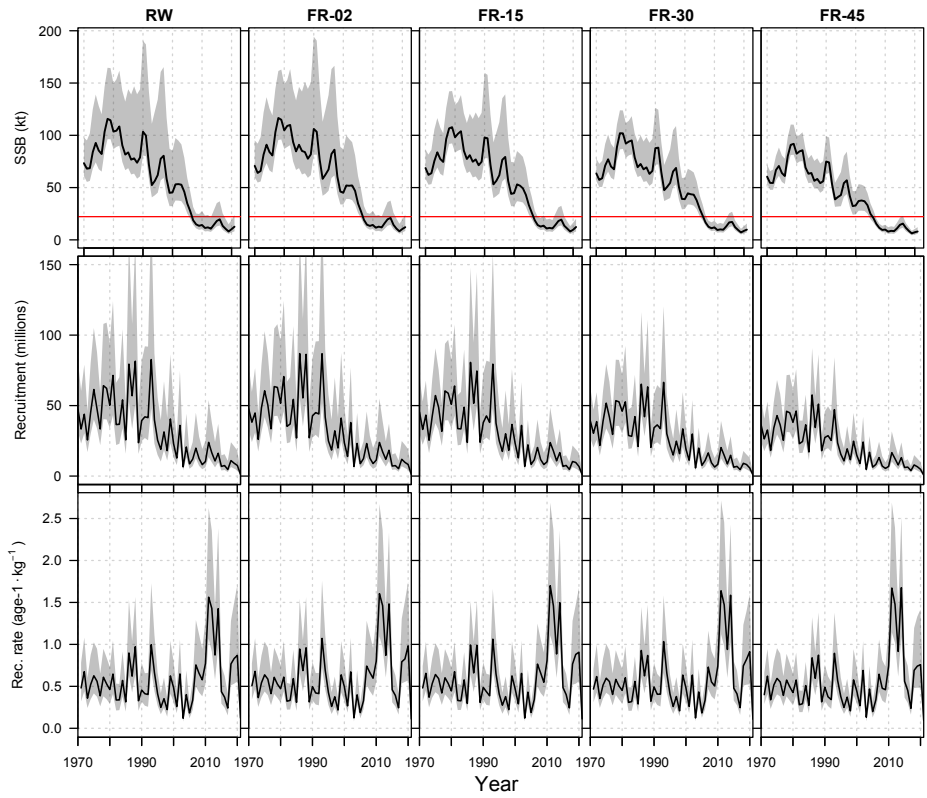

*Figure A11. Estimated stock-recruitment relationships for 4X5Y cod from 5 statistical catch-at-age models. A Beverton-Holt stock-recruitment function was fitted to each posterior sample of spawning stock biomass and the resulting recruitment one year later. Circles represent posterior modes of spawning biomass and recruitment, while lines and shaded regions represent medians and central 95% uncertainty intervals, respectively, of the fitted Beverton-Holt functions. The a and b estimates in the bottom right corner (posterior mode and central 95% uncertainty interval) represent the fecundity and density-dependent parameters of the Beverton-Holt function, respectively.*

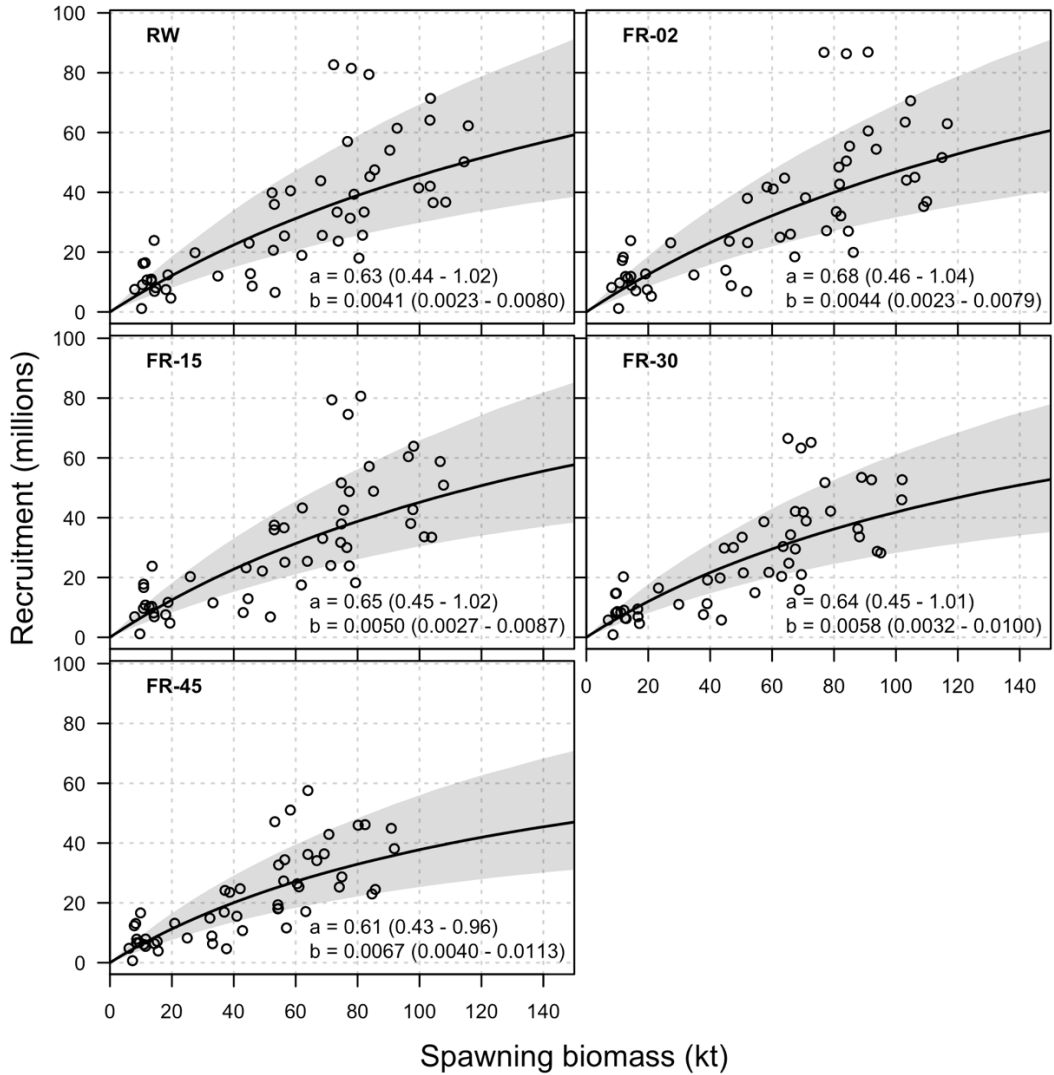

*Figure A12. Residuals (log(predicted) – log(observed)) for discard age-composition from the lobster fishery from the FR-15 model.*

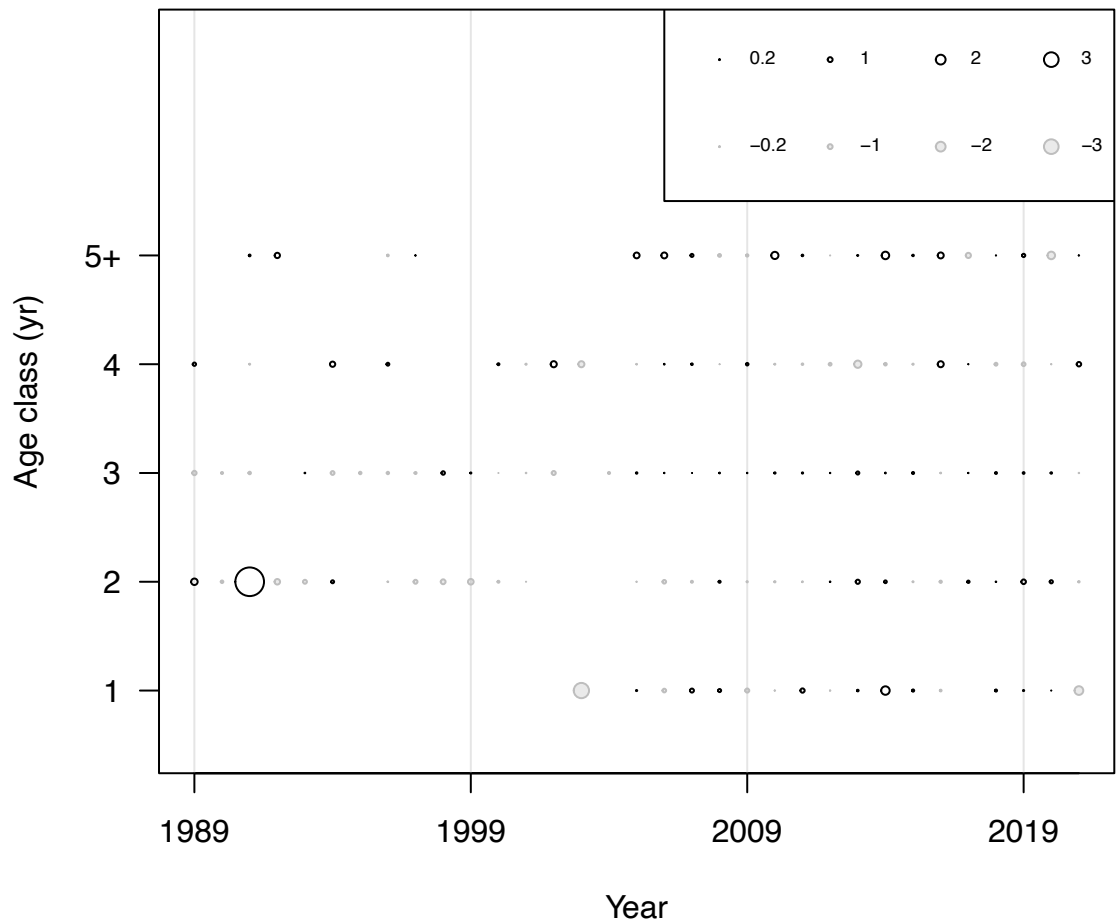

95  *Figure A13. Estimated selectivity-at-age of cod to the lobster fishery from the FR-15 model.*

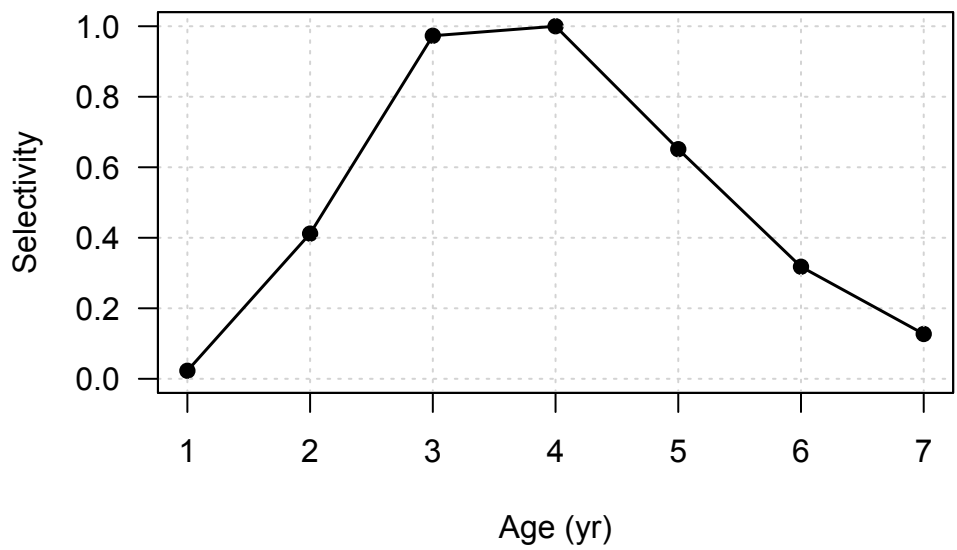

96
97

# Appendix B

# Estimating grey seal abundance along the southwest coast of Nova Scotia

## B.1 Background

Grey seals (*Halichoerus grypus*) in the northwest Atlantic Ocean comprise a single population with major breeding colonies at Sable Island and in the Gulf of St. Lawrence, and smaller colonies occurring along the coasts of Nova Scotia and the northeastern United States. Inferring abundance at the smaller grey seal colonies is difficult as observations at these colonies have greater uncertainty, comprise relatively short time series and dynamics are presumed to be driven by emigration from larger adjacent colonies. Recent assessment models for grey seals in Canadian waters have aggregated Sable Island and coastal Nova Scotia seals into a single management unit, thereby avoiding direct inferences about dynamics at the smaller colonies (Hammill et al. 2017; Rossi et al. 2021). Inferring grey seal foraging effort in NAFO divisions 4X and 5Y requires a more detailed approach, as seals from colonies along southwest Nova Scotia (SWNS) are expected to spend significantly more time foraging in 4X and 5Y than seals from Sable Island given the close proximity of seals along coastal Nova Scotia to these fishing areas. We therefore attempt to directly infer grey seal abundance along SWNS.

## B.2 Methods

Observations of pup production at SWNS colonies (Mud Is., Round Is., Noddy Is., and Flat Is.,) were available for 2007, 2010, 2016, and 2021 (den Heyer et al. 2017, 2021). The first two observations were visual counts in late January of mostly weaned pups, with many pups thought to have dispersed given the few mothers left at the colonies. For our analysis, we inflated these counts by 50% to account for pups that were missed in the visual counts or had dispersed before the survey. The latter two observations were based on aerial photographic surveys in early January, which were corrected for the small proportion of seals born after the survey date.

125  We used a logistic population model to represent SWNS pup production, i.e.,

$$\hat{I}_t = \frac{KI_0}{I_0 + (K - I_0) \exp(-rt)}$$

127  where $\hat{I}_t$ is pup abundance in year $t$, $r$ is the population growth rate, and $K$ is the carrying
128  capacity. We initialized the model ($t=0$) in 1980 and assumed no pup production prior to 1980.
129  Given the sparsity of the data, we were unable to estimate all three model parameters ($r$, $K$, and
130  $I_0$), so we fixed $I_0$ at 10, which was chosen to represent a small initial value. We tested the
131  sensitivity of model estimates to two alternative values of $I_0$ (1 and 100),

132  We specified weakly informative prior distributions for $r$ and $K$. We assigned a N(5000,10000)
133  on $K$ to represent our view that carrying capacity at SWNS is probably less than in the Gulf of St.
134  Lawrence, where the largest observed pup production was approximately 16,400. We specified a
135  N(0.12,0.24) prior on $r$, which was centered at the maximum observed growth rate at Sable
136  Island. We note that $r$ is driven by immigration and therefore does not represent an intrinsic
137  population growth rate; therefore, $r$ in our model may exceed the rate observed at Sable Island.

138  Observed pup counts $I_t$ were assumed to be arise from log-normal distributions, i.e.,

$$\log(I_t) \sim N\left(\log(\hat{I}_t), \sigma_t^2\right)$$

140  We calculated the variance as $\sigma_t^2 = (CV_t^2 + 1)$ where $CV_t$ was the coefficient of variation
141  around $\hat{I}_t$. CVs from statistical analyses were only available for the 2017 and 2021 observations.
142  We assigned large CVs (0.5) to the 2007 and 2010 observations as these counts are highly
143  uncertain.

144  The statistical model described above was implemented using the Template Model Builder (tmb)
145  package (Kristensen et al. 2016) within R version 4.2.1 (R Core Team, 2022). Bayes posterior
146  distributions for parameters and predictive pup count distributions were generated via a
147  Hamiltonian Monte Carlo (HMC; Duane et al. 1987, Neal 2011) algorithm known as the no-U-
148  turn sampler (NUTS; Hoffman and Gelman 2014) in the tmbstan R package (Kristensen 2018).
149  We ran three chains for 5000 iterations each, discarding the first half of each chain as a warm-up.
150  We monitored convergence using the potential scale reduction factor on rank-normalized split chains ($\hat{R}$), where $\hat{R} > 1.01$ indicates problems with convergence (Vehtari 2020), as well as the effective sample size (ESS) of the rank-normalized draws, where an ESS of at least 300 (100 per chain) was considered acceptable. We also monitor HMC-specific and NUTS-specific diagnostics (number of divergent transitions of number of transitions that saturated the maximum tree depth, respectively).

## B.3 Results

We did not detect any issues with model convergence; all $\hat{R}$ values were less than 1.01 and all ESS values were sufficiently large (Table B2). Additionally, there were no divergent transitions and no transitions that saturated the maximum tree depth.

The logistic model closely fit the two most recent observations of pup production and overestimated the two earliest observations (Figure B1). The relatively poor fit to the early pup production observations was not concerning given the high degree of uncertainty associated with those data.

The estimated growth rate was 0.17 (95% uncertainty interval [UI]: 0.15-0.20), while carrying capacity was 3,264 (95% UI: 2,481-5,897) (Figure B2. Prior and posterior distributions for the growth rate (r) and carrying capacity (K) of grey seal pup production at SWNS colonies.Figure B2).

Scaling estimated pup production by ratio of pups to adults from the grey seal assessment model implies a total grey seal abundance in SWNS colonies in 2021 of approximately 9,800 (95% UI: 8,600-11,600), growing to 11,900 (95% UI: 9,400-18,100) in 2030.

We did not detect model convergence issues when refitting the logistic model with $I_0 = 1$; however, fitting the model with $I_0 = 100$ caused 3.3% of post-warmup transitions to diverge, and tail ESS for $r$ was too low, suggesting inference from this model may be unreliable. Each model produced similar estimates of abundance over the most recent decade but projected abundance diverged thereafter, with abundance continuing to increase rapidly under $I_0 = 100$ while levelling off under $I_0 = 1$ (Figure B3). The posterior for abundance estimated under $I_0 = 1$ was almost entirely contained in the abundance posterior under $I_0 = 10$ between 2015 and 2030.

## Tables

*Table B1. Pup counts at grey seal colonies along southwest Nova Scotia. * indicates that the CV was assigned and was not based on statistical analysis.*

| Year | Raw count | Adjusted count | CV |
|---|---|---|---|
| 2007 | 204 | 306 | 0.50* |
| 2010 | 417 | 626 | 0.50* |
| 2016 | 1,849 | 2,105 | 0.07 |
| 2021 | 2,246 | 2,420 | 0.08 |

*Table B2. MCMC diagnostics for each model parameter: the potential scale reduction factor on rank-normalized split chains (), the effective sample size of the rank-normalized draws (BulkESS), the minimum of the effective sample sizes of the 5% and 95% quantiles (TailESS), the number of divergent transitions (#DT), and the number of transitions that saturated the maximum tree depth (#SMT).*

| Parameter |  | BulkESS | TailESS | #DT | #SMT |
|---|---|---|---|---|---|
| $r$ | 1.00 | 1496 | 1755 | 0 | 0 |
| $k$ | 1.00 | 1478 | 1746 | 0 | 0 |

*Figure B1. Logistic model fit to adjusted counts of SWNS pup production. The thick black line is the posterior median and the shaded region is the central 95% uncertainty interval. The vertical dashed line delimits the estimation and projection periods.*

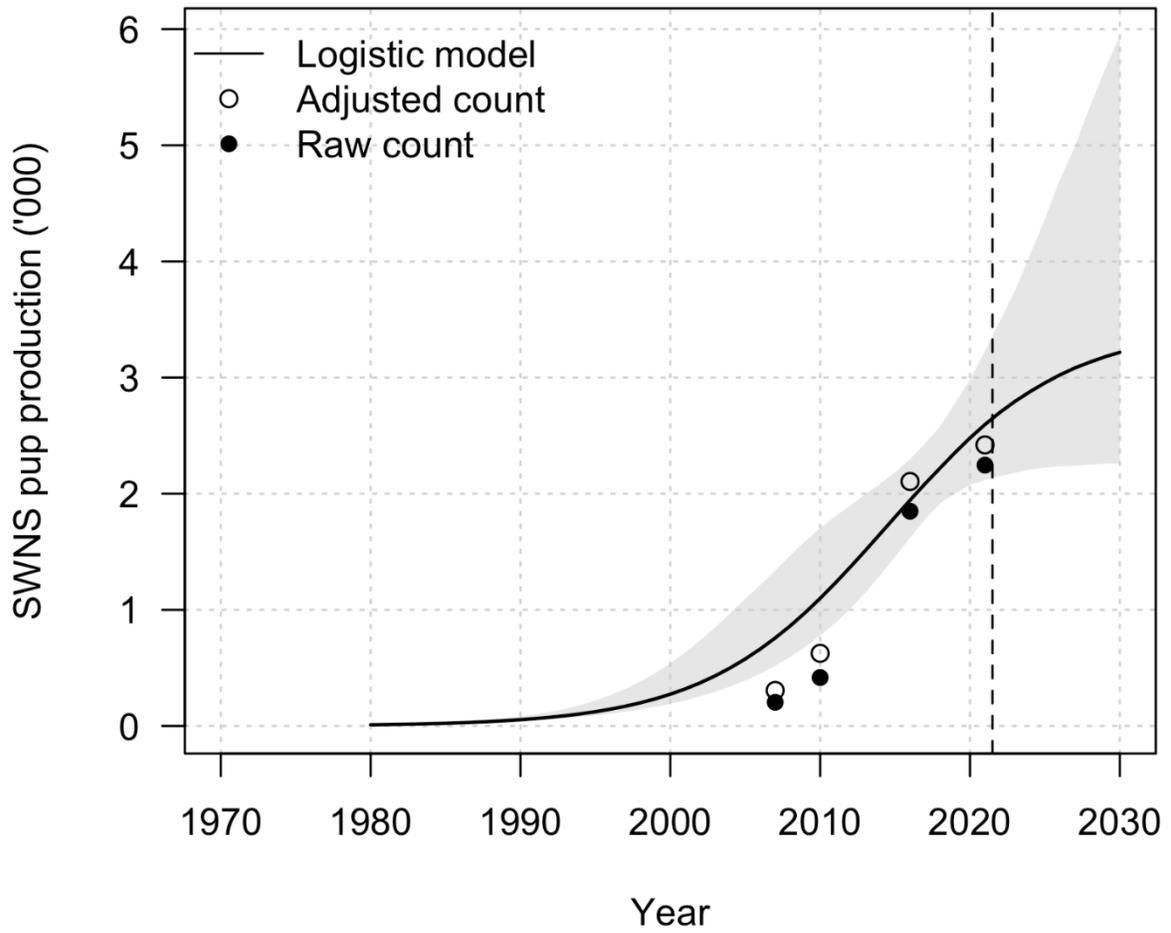

*Figure B2. Prior and posterior distributions for the growth rate (r) and carrying capacity (K) of grey seal pup production at SWNS colonies.*

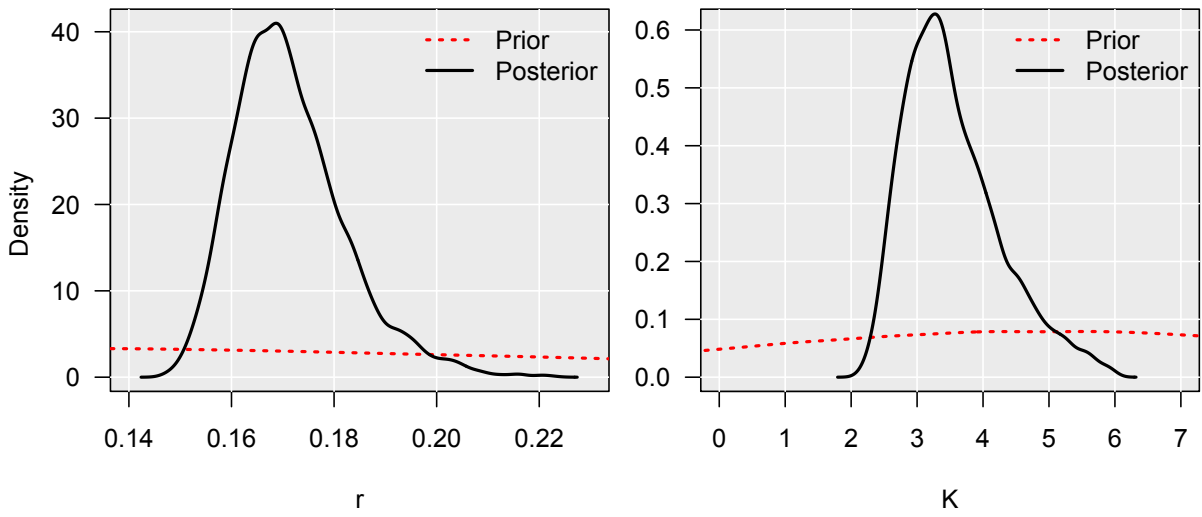

234  *Figure B3. Comparison of pup production (top row) and total abundance (bottom row) estimates for the*
235  *SWNS grey seal population from logistic models initialized with three values of $I_0$ (1, 10, and 100). Lines*
236  *are posterior medians and shaded regions are central 95% uncertainty intervals. The vertical dashed line*
237  *delimits the estimation and projection periods.*

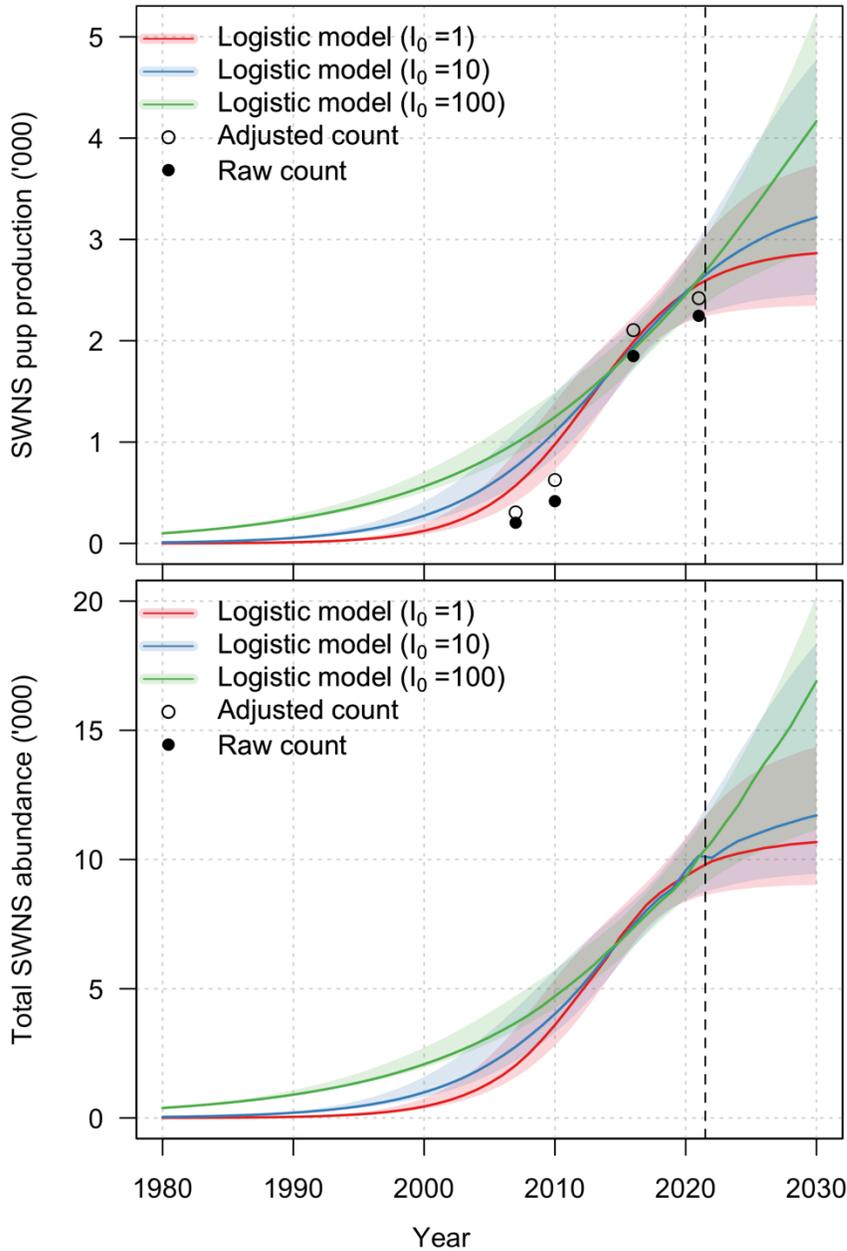

238